\documentclass[twocolumn,showpacs,amsmath,pra,aps,amssymb,superscriptaddress]{revtex4-1} 
\usepackage[T1]{fontenc}
\usepackage[latin9]{inputenc}
\usepackage{times}
\usepackage{color} 
\usepackage{xspace}
\usepackage{amssymb,amsmath}
\usepackage{amsbsy}
\usepackage[pdftex]{graphicx}
\usepackage{bm}
\usepackage{float}

\usepackage[unicode,breaklinks]{hyperref}
\hypersetup{
    unicode=true,
    plainpages=false, 
    colorlinks=true,
    linkcolor=blue,
    citecolor=blue,
    filecolor=black,
    urlcolor=blue
}
\usepackage{url}
\usepackage{verbatim}

\newcommand{\edd}{\epsilon_{\mathrm{dd}}}
 
\newcommand{\bk}{\mathbf{k}}
\newcommand{\br}{\mathbf{r}}

\begin{document}

\title{Ground-state phase diagram of a dipolar condensate with quantum fluctuations} 
\author{R.~N.~Bisset}   
\affiliation{INO-CNR BEC Center and Dipartimento di Fisica,
Universit\`a di Trento, Via Sommarive 14, I-38123 Povo, Italy}
\author{R.~M.~Wilson}   
\affiliation{Department of Physics, The United States Naval Academy, Annapolis, MD 21402, USA}
\author{D.~Baillie}  
\author{P.~B.~Blakie}   
\affiliation{Department of Physics, Centre for Quantum Science,
and Dodd-Walls Centre for Photonic and Quantum Technologies, University of Otago, Dunedin, New Zealand}

\begin{abstract} 
We consider the ground state properties of a trapped dipolar condensate under the influence of quantum fluctuations. 
We show that this system can undergo a phase transition from a low density condensate state to a high density droplet state, which is stabilized by quantum fluctuations. The energetically favored state depends on the geometry of the confining potential, the number of atoms and the two-body interactions.  We develop a simple variational ansatz and validate it against full numerical solutions. We produce a phase diagram for the system and present results relevant to current experiments with dysprosium and erbium condensates.  \end{abstract} 

\maketitle

\section{Introduction}
In Bose-Einstein condensates of dipolar (i.e.~highly magnetic) atoms, such as chromium \cite{Griesmaier2005a}, dysprosium \cite{Mingwu2011a} and erbium \cite{Aikawa2012a}, a dipole-dipole interaction (DDI)  between the atoms becomes important. The DDI is long ranged and anisotropic and gives rise to a rich array of new phenomena \cite{Lahaye_RepProgPhys_2009}. 
Many of these phenomena  (e.g.~roton-like excitations \cite{Santos2003a}, structured ground states \cite{Ronen2006b}) are predicted to occur only when the DDI is stronger than the short ranged $s$-wave interaction -- the so called dipole-dominated regime.    
However, in this regime the attractive component of the DDI (i.e.~head-to-tail attraction between dipoles) tends to destabilize the condensate making it susceptible to local or global collapse dynamics. The stability phase diagram, and collapse dynamics have received considerable experimental and theoretical attention \cite{Ronen2006a,Koch2008a,Ticknor2008a,Lahaye2009a,Parker2009a,Wilson2009a,Lu2010a,Muller2011a,Bisset2012,Corson2013a,Corson2013b,Linscott2014a}, and seemed to establish that the standard meanfield theory, i.e.~the Gross-Pitaevskii equation (GPE) with both  ($s$-wave) contact interaction and  non-local DDI terms, provides an accurate description of experiments in this regime.
 
However, recent experiments with dysprosium, aided by high resolution \textit{in situ} imaging have made new observations not accounted for by the standard meanfield theory \cite{Kadau2016a}. These experiments quenched a condensate into the dipole dominated regime and observed the formation of a stable droplet crystal. Each droplet contained $\sim\!10^3$ atoms, and had an estimated peak density about an order of magnitude larger than the pre-quenched condensate. In contrast to these observations, the meanfield theory predicts that the droplets would continue to collapse to extremely high densities where three-body recombination would cause rapid atomic loss and heating. Two suggestions have been made for the mechanism to stabilize these droplets at a finite size: (i) the presence of a conservative three-body interaction between the atoms \cite{Xi2016a,Bisset2015a,Blakie2016a} (also see \cite{Petrov2014a,Lu2015a}), and (ii) the role of beyond-meanfield quantum fluctuations \cite{Ferrier-Barbut2016a,Wachtler2016a,Saito2016a} (also see \cite{Petrov2015a}). Both effects can be accounted for by adding a new non-linear term to the standard meanfield theory with a higher order dependence on density than the usual two-body interactions. Due to this density dependence, these terms have a limited effect in the low density (pre-quenched) condensate, but can be significant in the high density droplets. Recent path-integral Monte Carlo calculations by Saito \cite{Saito2016a} have provided strong quantitative evidence that quantum fluctuations alone are able to stabilize droplets, without the need for a three-body interaction.

In this paper we investigate the role of quantum fluctuations on the ground state of a dipolar condensate. 
These fluctuations are accounted for at lowest order by corrections to the equation of state for a condensate of hard spheres, originally predicted by Lee, Huang, and Yang (LHY) \cite{LY1957,LHY1957}. The extension of these LHY corrections to the case of a condensate with DDIs was developed in Refs.~\cite{Schatzhold2006a,Lima2011a,Lima2012a}.  At normal atomic condensate densities ($\!\sim\!10^{20}\,$m$^{-3}$) these  LHY corrections are negligible. For this reason experiments with non-dipolar atoms have worked with strongly interacting atoms (i.e.~using resonances to enhance the scattering length) to measure LHY corrections (e.g.~see \cite{Altmeyer2007a,Papp2008a,Navon2011a}).
In contrast, the LHY corrections become important for the droplets, which achieve much higher densities. 

In order to investigate the nature of the ground states we use a generalized GPE, in which the meanfield theory is augmented with a local density treatment of the quantum fluctuations. The accuracy of this approach, even when a high density droplet-type ground state emerges, has been established in Ref.~\cite{Saito2016a}, albeit for small atom numbers ($\sim\!10^3$) where path-integral Monte Carlo calculations were feasible. We use two approaches to calculate the ground states of this formalism: (1) a simple variational treatment and (2) full numerical solutions of the generalized GPE. We validate that both approaches are in good qualitative agreement over a wide parameter regime.  We find that this theory yields two types of ground states, dependent on the system parameters such as trap shape, condensate number and the $s$-wave scattering length. The first type of state, which we refer to as the low density phase (LDP),  is the usual type of stable condensate observed in experiments. The density profile of LDPs is dominated by the interplay of the two-body interactions and the trapping potential (cf.~Thomas-Fermi solution \cite{Dalfovo1999}). The  second type of state, referred to as the high density phase (HDP), is a single high density droplet that forms due to the attractive character of the DDI. The quantum fluctuations play a crucial role in stabilizing the droplet.
We find that there are two different regimes of the HDP depending whether kinetic energy (quantum pressure) plays a significant role which, in turn, depends on the atom number.
In our results we make predictions for both $^{164}$Dy and $^{168}$Er, and find that both species should be suitable to explore our predicted phases and phase transitions.

\section{Formalism}

\subsection{Generalized GPE theory}
Our system of interest is a harmonically trapped dipolar condensate.  We consider a cylindrically symmetric geometry, i.e.~the confining potential is cylindrically symmetric about the $z$ axis, and the dipoles are polarized along $z$. This choice is a good approximation to the motivating experiments reported in Ref.~\cite{Kadau2016a}, and affords a more efficient and accurate solution for the ground states.

Within a local density treatment of the quantum fluctuations we can introduce a  generalized time-dependent Gross-Pitaevskii equation \footnote{The dipolar LHY local density treatment was formulated in \cite{Lima2011a,Lima2012a} and was recently applied as a generalized GPE in  \cite{Wachtler2016a,Saito2016a}}  \begin{align}
i\hbar\frac{\partial \psi}{\partial t}&= \mathcal{L}_{\mathrm{GP}}\psi,\\
=&\left[H_{\mathrm{sp}}\!+\!\int\!d\br'\,U(\br\!-\!\br')|\psi(\br')|^2+\gamma_{\mathrm{QF}}|\psi|^3\right]\psi,\ \label{GPE}
\end{align}
where 
\begin{align}
H_{\mathrm{sp}}=-\frac{\hbar^2\nabla^2}{2m}+\frac{1}{2}m(\omega_\rho^2 \rho^2+\omega_z^2z^2),
\end{align}
is the single particle Hamiltonian including a harmonic confinement potential with trapping frequencies $\{\omega_\rho,\omega_z\}$, where $\rho=\sqrt{x^2+y^2}$ is the radial coordinate.
The two-body interactions between atoms are described by the pseudo-potential \cite{Lahaye_RepProgPhys_2009}
\begin{align}
U(\mathbf{r})=g\delta(\br)+\frac{\mu_0\mu^2}{4\pi}\frac{1-3\cos^2\theta}{r^3},
\end{align}
where  $g={4\pi a_s\hbar^2}/{m}$, with $a_s$ being the $s$-wave scattering length. The long-ranged DDI term is for dipoles of magnetic moment $\mu$ polarized along $z$  with $\theta$ being the angle between $\br$ and the $z$-axis.

\begin{figure}[htbp]
   \centering
  \includegraphics[width=3.1in]{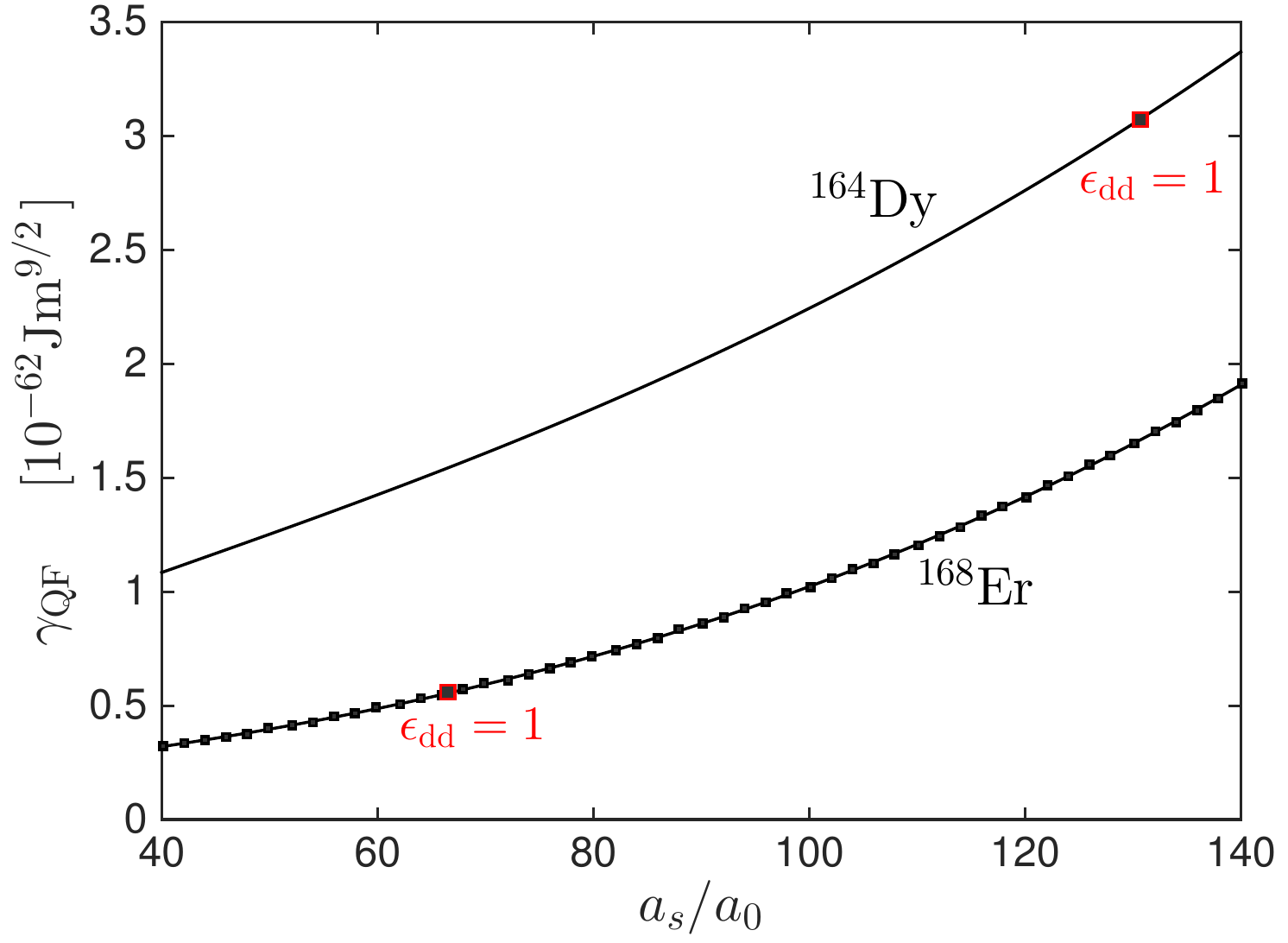} 
  \vspace*{-0.1cm}
   \caption{(color online)   The quantum fluctuation parameter $\gamma_{\mathrm{QF}}$ for each species is indicated for (solid line) $^{164}$Dy  and (solid line with symbols) $^{168}$Er cases as a function of $a_s$ (where $a_0$ is the Bohr radius). The values where $\epsilon_{\mathrm{dd}}=1$ for each species are indicated by small red boxes. }
   \label{FigQF}
\end{figure} 

 The last term in  Eq.~(\ref{GPE})  accounts for the quantum fluctuations. In a homogeneous dipolar condensate quantum fluctuations are predicted to  shift the chemical potential, a correction of  the form $ \Delta\mu=\gamma_{\mathrm{QF}}n^{3/2}$  \cite{Lima2011a}, where $n$ is the density. The quantum fluctuation parameter $\gamma_{\mathrm{QF}}$ is determined by the excitation spectrum, and thus depends on both the contact and DDIs. Making the local density approximation by setting $n \to n(\mathbf{r}) = |\psi(\mathbf{r})|^2$ yields the term appearing in our generalized GPE.
Some evidence for the applicability of the fluctuation term [as used in Eq.~(\ref{GPE})] in the regime of (HDP) droplet ground states  has been provided
by recent path-integral Monte Carlo calculations \cite{Saito2016a} for cases with $\sim10^3$ atoms.  More such studies, particularly at larger numbers are necessary to assess the accuracy of this treatment over a broader parameter regime relevant to experiments. The general validity requirement for including quantum fluctuations in the form we use here  is that the gas parameter is small (or equivalently that the depletion remains small, see Ref.~\cite{Wachtler2016a}). A separate concern is the applicability of the local density approximation.
Previous works concerning the excitations of   trapped dipolar condensates have demonstrated that the local density approximation works well \cite{Blakie2012a,Bisset2013a,JonaLasinio2013} in the regime of LDP ground states, although these results were typically in the regime of reasonably large condensates where the density varies slowly. It is possible to include quantum fluctuation effects without making the local density approximation (i.e.~explicitly diagonalizing for the quasi-particle modes), as has been done for condensates with contact interactions (e.g.~see \cite{Hutchinson1998a,Castin1998a,Morgan2000a,Morgan2004a}). For the case of DDIs this task is more technically challenging because exchange interactions involve computing the two-point correlation function of the depleted atoms (e.g.~see \cite{Cormack2012a,Ticknor2012a}).

The quantum fluctuation parameter has the analytic form
\begin{align}
\gamma_{\mathrm{QF}}=\frac{32}{3}g\sqrt{\frac{a_s^3}{\pi}}\left(1+\frac{3}{2}\epsilon_{\mathrm{dd}}^2\right),\label{gammaV}
\end{align}
 where the dimensionless number $\epsilon_{\mathrm{dd}}=a_{\mathrm{dd}}/a_s$, quantifies the relative strength of the DDI and the $s$-wave interaction, where $a_{\mathrm{dd}}=m\mu_0\mu^2/12\pi\hbar^2$ is the \textit{dipole length}. When $\epsilon_{\mathrm{dd}}>1$ we refer to the system as being dipole dominated, and it was in this regime that the droplet crystal was observed to form in experiments. 
Details of how we obtain result (\ref{gammaV}) from the Bogoliubov theory of a  homogeneous dipolar condensate, and a brief comparison to the other approaches used in the literature, are given in the Appendix.
 
 It is interesting to quantify the size of $\gamma_{\mathrm{QF}}$, and hence quantum fluctuations, for $^{164}$Dy and $^{168}$Er atoms.  We show results for both atoms in  Fig.~\ref{FigQF}  as a function of $a_s$. On each curve the location where $\epsilon_{\mathrm{dd}}=1$ is indicated, noting that the dipole dominated regime lies to the left of this point. We observe that at $\epsilon_{\mathrm{dd}}=1$  the value of $\gamma_{\mathrm{QF}}$ for $^{164}$Dy is approximately six times larger than the value for $^{168}$Er, suggesting that at a given density and at $\epsilon_{\mathrm{dd}}\approx1$, the role of quantum fluctuations will be much larger for dysprosium.

\subsection{Full numerical solution for stationary states} 
It is of interest to find stationary states of the generalized GPE (\ref{GPE}) in order to characterize equilibrium or meta-stable states of the system. These states can be obtained by numerically solving the time-independent GPE
\begin{align}
\mathcal{L}_{\mathrm{GP}}\psi_0=\mu_c\psi_0,\label{tiGPE}
\end{align}
where $\psi_0$ is the stationary solution wavefunction, taken to be normalized to the number of atoms $N$, and $\mu_c$ is the condensate chemical potential. Because the system is cylindrically symmetric we can extend the  algorithm developed in Ref.~\cite{Ronen2006b} to solve for these states. We emphasize that this algorithm allows us to follow stationary states even once they become meta-stable (i.e.~are no longer the global ground state).

\subsection{Variational solution}
A simpler alternative to finding full numerical solutions of the time-independent generalized GPE is to consider a variational approach.
Here we do this by approximating the condensate by the Gaussian ansatz
\begin{equation}
\psi_{\mathrm{var}}=\sqrt{\frac{8N}{\pi^{3/2}\sigma_\rho^{2}\sigma_z }}\exp\left[-2\left(\frac{\rho^{2}}{\sigma_\rho^{2}}+\frac{z^{2}}{\sigma_z^{2}}\right)\right],\label{psivar}
\end{equation}
where $\sigma_\rho$ and $\sigma_z$ are the full widths at 1/$e$ of the peak density along $\rho$ and $z$, respectively.
In order to variationally determine the stationary state parameters we evaluate the energy functional corresponding to the generalized GPE (\ref{tiGPE}), i.e.~
\begin{align}
&E[\psi] = \label{Efuncfull} \\
&\int d\br\,\psi^*\Big[H_{\mathrm{sp}}\!+ \frac{1}{2} \!\int\!d\br'\,U(\br\!-\!\br')|\psi(\br')|^2 +\frac{2}{5}\gamma_{\mathrm{QF}}|\psi|^3  \Big]\psi, \nonumber
\end{align}
with the last term corresponding to the energy correction due to quantum fluctuations within the local density approximation [cf.~$\Delta E$ in Eq.~(\ref{e:DEfull})].
Evaluating the energy functional (\ref{Efuncfull}) using the variational ansatz yields
\begin{align}
  \frac{E(\sigma_\rho,\sigma_z)}{N\hbar\omega_z}=& {\left(\frac{2a_z^2}{\sigma_\rho^{2}}+\frac{a_z^2}{\sigma_z^{2}}\right)} + \left(\frac{1}{8\lambda^2}\frac{\sigma_\rho^{2}}{a_z^2}+\frac{\sigma_z^{2}}{16a_z^2}\right) 
   \label{EfuncG} \\
  &+{\frac{8Na_z^3}{\sqrt{2\pi}\sigma_\rho^{2}\sigma_z}\left[\frac{a_s}{a_z}-\frac{a_{\mathrm{dd}}}{a_z}f\left(\frac{\sigma_\rho}{\sigma_z}\right)\right]} \nonumber \\
  &+ {\frac{128N^{3/2}\gamma_{\mathrm{QF}}}{25\sqrt{5}\pi^{9/4}\sigma_\rho^{3}\sigma_z^{3/2} \hbar\omega_z}}\nonumber,
 \end{align} 
 where
 \begin{align}
 f(x)=\frac{1+2x^2}{1-x^2}-\frac{3x^2\mathrm{arctanh}\sqrt{1-x^2}}{(1-x^2)^{3/2}}
 \end{align}
and $a_z=\sqrt{\hbar/m\omega_z}$.
We identify the stationary solutions by numerically locating the values of $\sigma_\rho$ and $\sigma_z$ that minimize the energy functional \eqref{EfuncG}.

\section{Results}
\subsection{Ground and metastable states}
   \begin{figure}[htbp]
   \centering
  \includegraphics[width=3.4in]{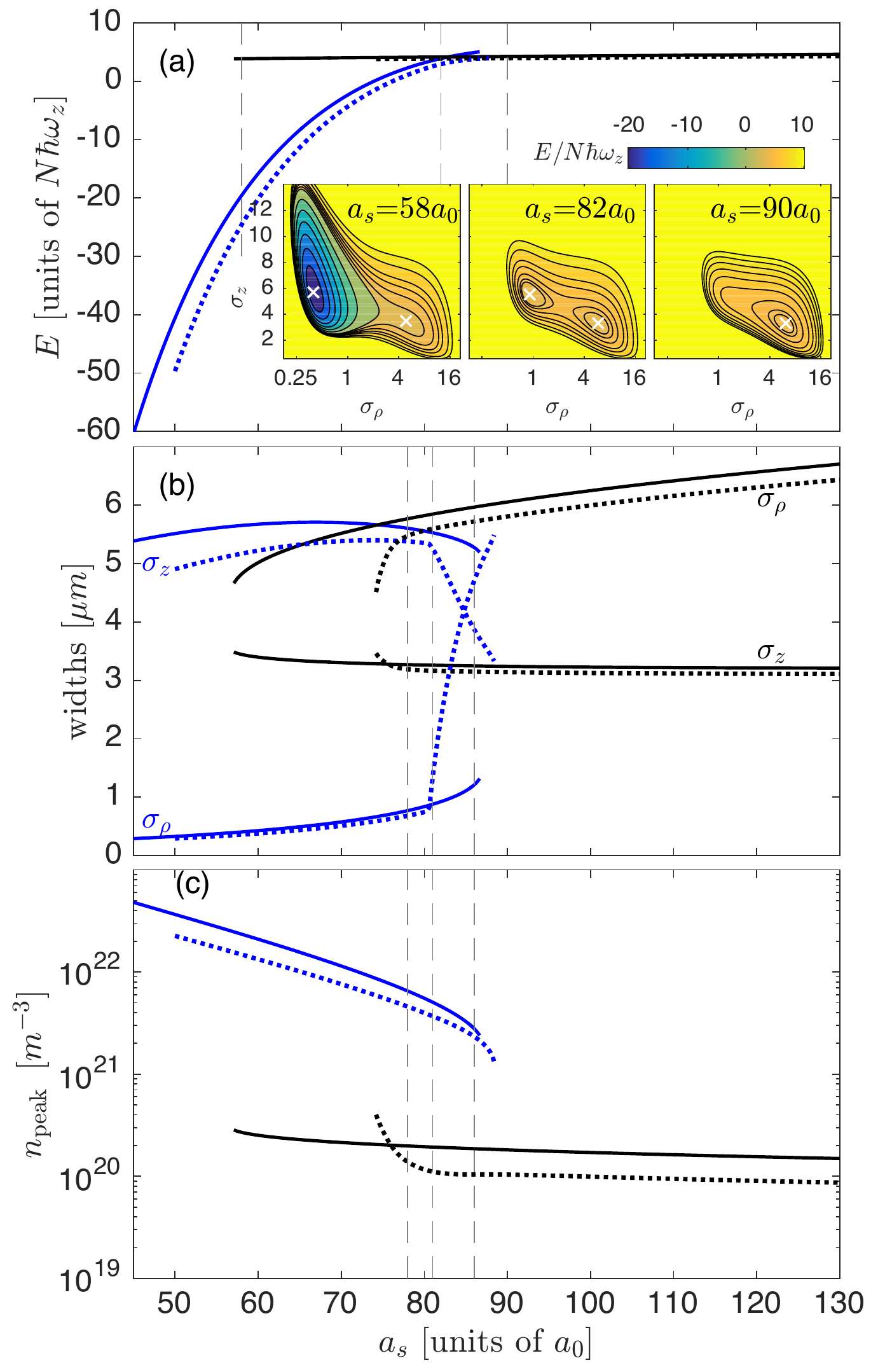}
   \caption{ Stationary solution properties in an oblate trap as a function of the $s$-wave scattering length.  (a) Energy, (b) width parameters, and (c) peak density. The variational and GPE  solutions are given as solid and dotted lines respectively. Insets to (a) show contours of $E(\sigma_\rho,\sigma_z)$ highlighting the minima, for the $a_s$ values indicated [also shown as  vertical dashed lines in (a)].  Vertical dashed lines in (b) and (c) represent the cases studied in Fig.~\ref{FigCondDen}.
Results are for the case of $N=15\times10^3$ $^{164}$Dy atoms with $a_{\mathrm{dd}}=130\,a_0$,  where $a_0$ is the Bohr radius, and $\{\omega_\rho,\omega_z\}=2\pi\times\{45,133\}\,$s$^{-1}$.
}
   \label{FigBranches}
\end{figure} 
 We first consider the nature of stationary solutions for the parameter regime of the experiments of Kadau \textit{et al.}~\cite{Kadau2016a}.  Results for the stationary state properties (from both the variational solutions and the full numerical solutions) are shown in Fig.~\ref{FigBranches} as a function of the $s$-wave scattering length.  
 
 The energies of these solutions [Fig.~\ref{FigBranches}(a)] reveal two distinct energy branches within the range of scattering lengths considered. The upper energy branch corresponds to a solution of low peak density [see Fig.~\ref{FigBranches}(c)], and we label it as the low density phase (LDP). This branch is the ground state for $a_s\gtrsim85\,a_0$. For values of $a_s$ below this the other branch is the ground state, as it rapidly decreases in energy with decreasing $a_s$. The corresponding solutions have a high peak density [see Fig.~\ref{FigBranches}(c)] and we label this branch as the high-density phase (HDP).  The insets to Fig.~\ref{FigBranches}(a) show the variational energy surface as a function of $(\sigma_\rho,\sigma_z)$, revealing the two solutions as local minima.  
 
\begin{figure}[tbp]
   \centering
 \includegraphics[width=3.3in]{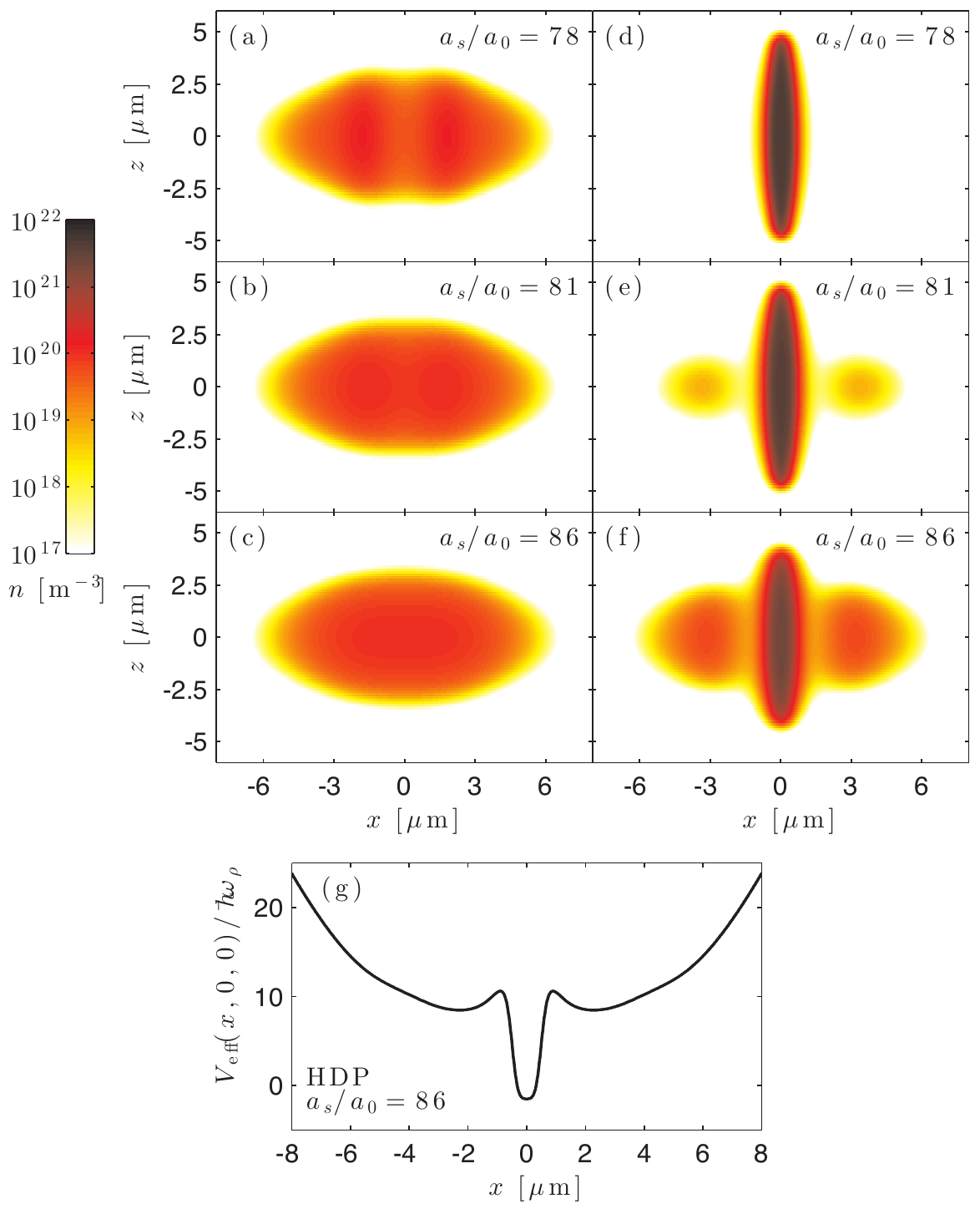}
   \caption{Density slices of stationary state solutions to the generalized GPE in the $y=0$ plane for the LDP (a)-(c) and the HDP (d)-(f) for three regimes $a_s/a_0 = 78$, 81 and 86  indicated as vertical dashed lines in Figs.~\ref{FigBranches} (b) and (c). For $a_s/a_0=78$ the LDP exhibits a `blood-cell' profile, while for $a_s/a_0=81$ and 86 the HDP has a Saturn ring-like feature. (g) The effective potential as defined in Eq.~(\ref{Eq:Veff}) for the case of $a_s/a_0=86$. }
   \label{FigCondDen}
\end{figure}

The widths of the solutions are shown in Fig.~\ref{FigBranches}(b).   For the full GPE solutions we extract the widths using the moments 
  \begin{align}
 \sigma_\rho^2 &=\frac{4}{N}\int d\mathbf{r}\,\rho^2|\psi(\mathbf{r})|^2,\quad
 \sigma_z^2 =\frac{8}{N}\int d\mathbf{r}\,z^2|\psi(\mathbf{r})|^2,
 \end{align}
 chosen  to correspond to the variational width parameters when applied to the variational state (\ref{psivar}).
  The LDP solution has an oblate geometry with an aspect ratio similar to that of the trap, although distorted slightly by the anisotropic character of the DDI (this is the typical  behavior of the dipolar Thomas-Fermi solution \cite{Eberlein2005a}). In contrast the HDP solution is a narrow prolate droplet with $\sigma_\rho\ll \sigma_z$. This configuration reduces the dipolar energy, but has an appreciable increase in the $z$-extent of the condensate and hence the $z$-component of the trap  potential energy increases. We notice in Fig.~\ref{FigBranches}(b) that the HDP widths obtained from the full GPE solution have a sudden kink and change in behavior for $a_s\gtrsim80\,a_0$. This happens as the solution develops a radial lobe outside the main condensate, a feature that is not captured by the variational ansatz. Examples of these condensate profiles are shown in Fig.~\ref{FigCondDen}. This lobe occurs because the effective potential for the atoms in the condensate
 \begin{align}
 V_{\mathrm{eff}}(\mathbf{r})=V_{\mathrm{trap}} +\frac{1}{2}\int\!d\mathbf{r}'U(\mathbf{r}-\mathbf{r}')|\psi_0(\mathbf{r}^\prime)|^2+\frac{2\gamma_{\mathrm{QF}}}{5}|\psi_0|^3, \label{Eq:Veff}
 \end{align}
  [i.e.~the non-kinetic part of the integrand in Eq.~(\ref{Efuncfull}), with $V_{\mathrm{trap}}$ denoting the harmonic trap] has a local minimum outside the condensate, and an appreciable number of atoms can tunnel into this (cf.~Saturn ring instability discussed in 
 Ref.~\cite{Eberlein2005a}).
 
 The peak density $n_{\mathrm{peak}}$ of the stationary solutions is shown in Fig.~\ref{FigBranches}(c). This is taken to be $|\psi_{\mathrm{var}}(\mathbf{0})|^2$ and $\max\left\{|\psi_c|^2\right\}$ for the variational and full GPE solutions \footnote{In some regimes the full GPE solutions can have peak density occurring away from trap center (e.g.~see \cite{Ronen2006a,Lu2010a,Martin2012a}).}, respectively. The peak density results emphasize the significant quantitative change in density between the two branches, i.e.~the HDP branch is approximately two orders of magnitude more dense than the LDP branch.

\begin{figure}[htbp]
   \centering
  \includegraphics[width=3.4in]{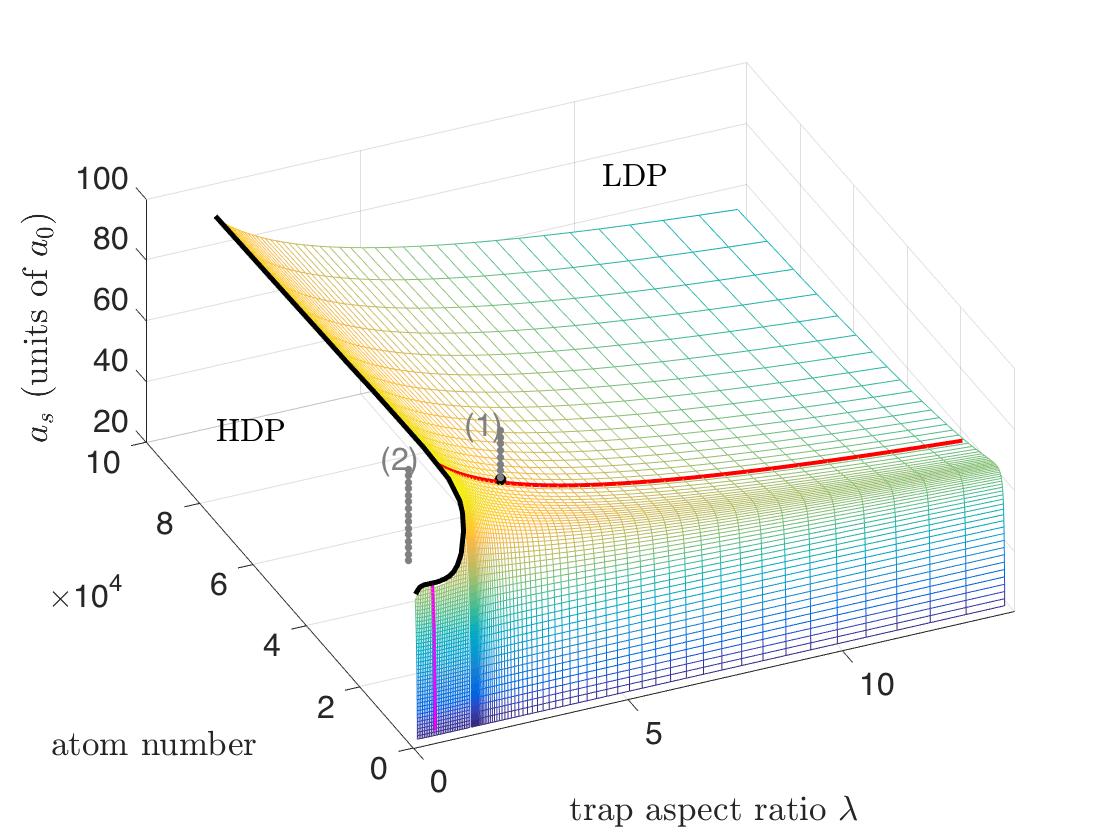}
  \vspace{-0.5cm}
   \caption{(color online) Surface marking the phase transition between the low density (LDP) and high density (HDP) phases of a dysprosium dipolar condensate. The critical line is indicated as a thick black line. Parameters are for $^{164}$Dy, and the harmonic trap is taken to have a fixed geometric mean trap frequency of  $\bar\omega/2\pi=64.6\,$Hz. Phase diagram calculated using the variational ansatz. Examples of two paths are indicated as vertical lines that correspond  to an $s$-wave quench that (1) passes across the phase transition, and  (2) avoids the phase transition by evolving continuously from the LDP to HDP  outside the critical line.  Path (1) is the interaction quench used in Ref.~\cite{Kadau2016a}. The two curves which lie on the surface are the cuts that are explored further in Figs.~\ref{Fig:PT_bigN} (a) and \ref{Fig:PDsmallN} (c).  }
   \label{FigQFPD}
\end{figure} 
  
\subsection{Phase diagram} \label{Sec:CPgeom}

Having validated that the variational approach is reasonably accurate, we can employ it to calculate a phase diagram over a wide parameter regime for the system. We find that the interesting parameters to explore are the condensate number, $s$-wave scattering length and trap aspect ratio $\lambda=\omega_z/\omega_\rho$. The trap aspect ratio is important because it influences the condensate geometry which directly affects the DDI energy. That is, by virtue of the long-range and anisotropic character of the DDI, its average energy changes from positive to negative as the condensate shape changes from oblate to prolate.
In general we can identify whether a stationary solution is in the LDP or the HDP by the distinctive peak densities associated with these phases. In this way we are able to make a phase diagram in $(N,\lambda,a_s)$-space. 
We show the result of such a phase diagram calculated for   $^{164}$Dy in Fig.~\ref{FigQFPD}. This figure shows the phase transition surface, where the energies of the LDP and HDP are degenerate
 [e.g.~the intercept of the two branches at $a_s\approx82\,a_0$ in Fig.~\ref{FigBranches}(a)]. Above this surface (i.e.~for higher values of $a_s$) the LDP is the ground state, while below this surface the HDP is the ground state. This phase transition surface terminates on a critical line, and for aspect ratios $\lambda\lesssim1$ and $N\gtrsim10^3$ there is no phase transition between the LDP and HDP, but a continuous evolution between these two phases. This behavior is similar to the usual gas-liquid phase transition, which also involves two states primarily differing by density. In that case it is also possible to  pass between the liquid and gas phases continuously by going around the critical point.
 
From Fig.~\ref{FigQFPD} we also observe that there are two distinct regimes of behavior for the phase diagram: At large $N$ ($\gtrsim5\times10^3$) the phase transition surface is almost independent of $N$, while at small $N$ the surface decreases rapidly with decreasing  $N$. We now explore these two regimes in more detail.
 
     \begin{figure}[tbp]
   \centering
  \includegraphics[width=3.2in]{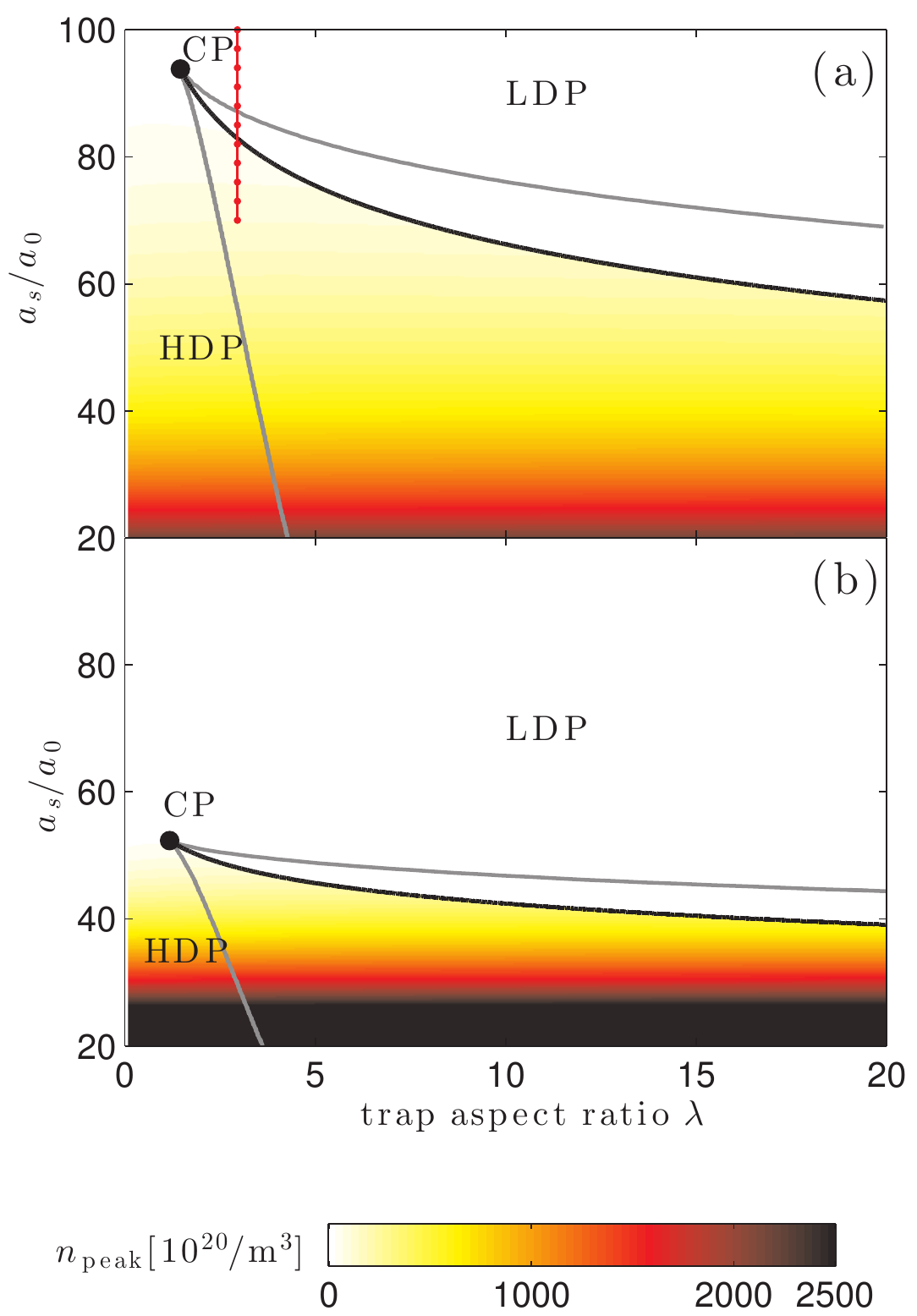}
  \vspace*{-0.2cm}
   \caption{(color online) Phase diagram as a function of $s$-wave scattering length and trap aspect ratio for (a) dysprosium and (b) erbium. The shading represents the peak density of the ground state. Black line indicates the phase transition, the large dot marks the critical point (CP) and the gray lines bound the region of metastable coexistence. Note that for erbium the density eventually exceeds the maximum given in the colorbar but we allow this saturation to facilitate the comparison with dysprosium. The HDP generally exists at smaller $a_s$ than the LDP. Parameters: $N=15000$, $\bar\omega/2\pi = 64.6$ Hz. The red line with dots indicates the quench path used in Ref.~\cite{Kadau2016a}.
   }
   \label{Fig:PT_bigN}
\end{figure} 

\subsubsection{Large-$N$ regime}\label{Sec:largeN}
 In Fig.~\ref{Fig:PT_bigN}(a) we show a slice of the phase diagram in ($\lambda,a_s$)-space for $N=15\times10^3$ $^{164}$Dy atoms. We have shaded this plot by the ground state peak density to reveal the sudden change that occurs upon crossing the phase boundary.
As the critical point is approached along the phase transition line, the difference in the HDP and LDP densities decreases.
 For reference, the phase transition line in this figure corresponds to the red line on the phase boundary surface in Fig.~\ref{FigQFPD}.
  In this large-$N$ regime the phase transition only occurs in oblate traps, arising from an interplay of the anisotropy of the DDI and the harmonic confinement. The stationary states for this case, with $\lambda=2.96$, were already presented in Fig.~\ref{FigBranches}, revealing the typical properties of the LDP and HDP states in this regime.
  Notably, in the LDP the trap potential energy is minimized by the condensate (approximately) adopting the aspect ratio of the trap, and in this configuration the DDI energy is positive. In the HDP the condensate adopts a prolate geometry that minimizes the DDI, but at the expense of higher contact-interaction, quantum-fluctuation and $z$-component of trap potential energies.
 Generally we observe that as $a_s$ decreases the ground state density increases. This is because below the phase boundary the quantum fluctuations play an important role in stabilizing the condensate against the attractive DDIs. As $a_s$ and hence the value of $\gamma_{\mathrm{QF}}$ decreases (see Fig.~\ref{FigQF}), the peak density has to increase for the quantum fluctuation term to balance the DDIs.

   In Fig.~\ref{Fig:PT_bigN}(b), we show the phase diagram for $^{168}$Er. Because this atom has a much smaller value of $a_{\mathrm{dd}}$ the dipole dominated regime occurs at a much lower $s$-wave scattering length, hence the phase boundary is at lower values of $a_s$ compared to the dysprosium results. At these values of $a_s$ the value of $\gamma_{\mathrm{QF}}$ is much smaller than for $^{164}$Dy (see Fig.~\ref{FigQF}) and hence the peak density of the droplets is much higher.

      \begin{figure*}[htbp]
   \centering
  \includegraphics[width=5.5in]{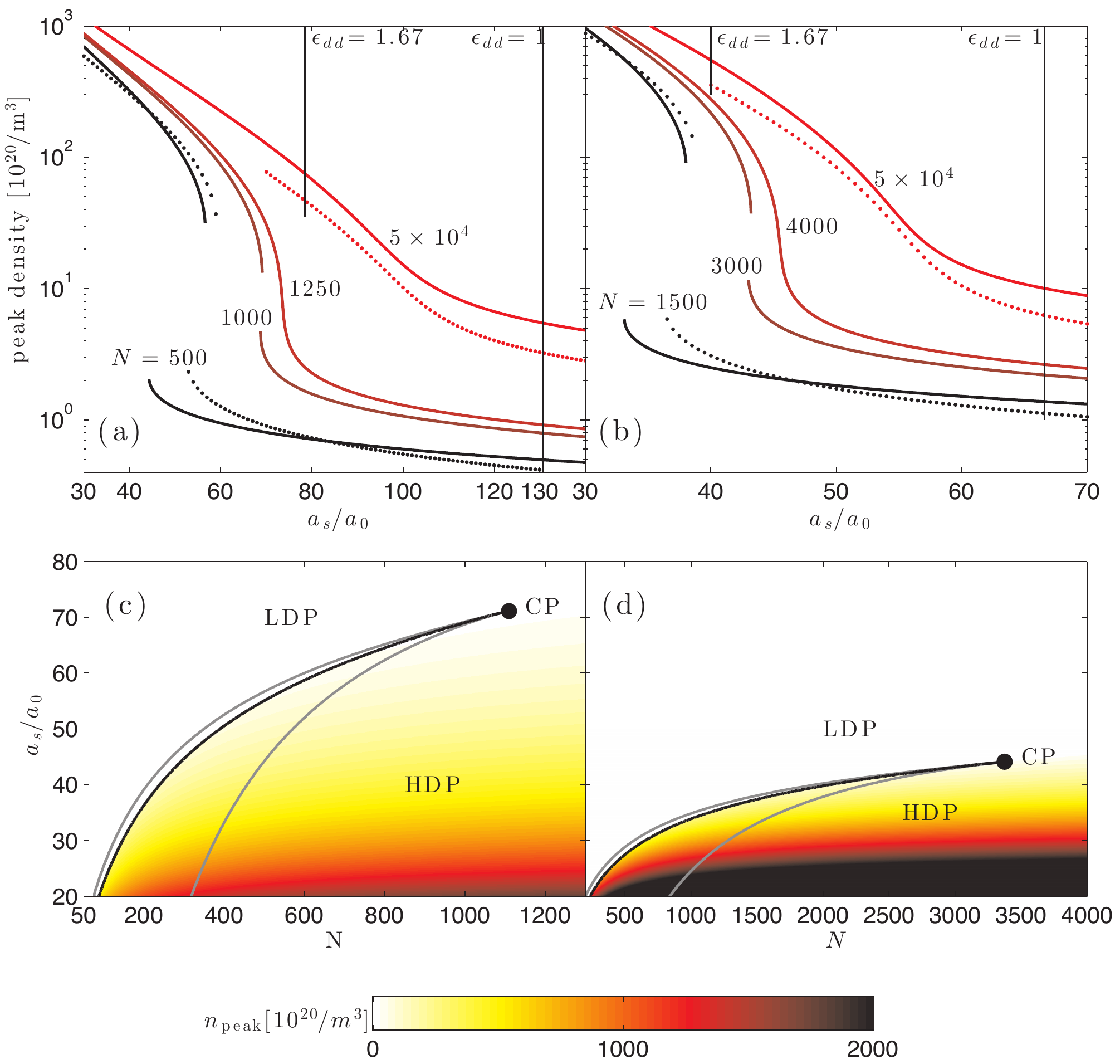}
  \vspace*{-0.05cm}
   \caption{Upper: peak density versus scattering length.
   (a) Dysprosium atom numbers from black to red (gray): 500, 1000, 1250, $5\times10^4$.
   (b) Erbium atom numbers from black to red (gray): 1500, 3000, 4000, $5\times10^4$. 
 Solid curves represent variational solutions while points are from the full GPE, the latter of which is only shown for the smallest and largest atom numbers. The vertical solid lines illustrate $\epsilon_{\mathrm{dd}}=1$ and $\epsilon_{\mathrm{dd}}=0.6$.
 Lower:
 (c) Dysprosium phase diagram in $a_s$-$N$ space demonstrating a critical point at $N$ = 1110, $a_s/a_0$ = 71.1.
   (d) Erbium critical point at $N$ = 3380, $a_s/a_0$ = 44.1.  The shading represents the peak density of the ground state. The black lines indicate the phase transition, the large dots mark the critical points and the gray lines bound the region of metastable coexistence. Note that for erbium the density eventually exceeds the maximum given in the colorbar but we allow this saturation to facilitate the comparison with dysprosium.  Trap parameters are the same for all panels: $\omega_\rho=2\pi\times 81.4$ s$^{-1}$, $\omega_z=2\pi\times 40.7$ s$^{-1}$.
   }
   \label{Fig:PDsmallN}
\end{figure*} 
\subsubsection{Small $N$ regime}
We now turn to considering the behavior of the phase diagram in the small-$N$ regime. Here the kinetic energy (quantum pressure) term becomes important in determining the stationary state properties. We find that a phase transition between the LDP and HDP can occur for all trap aspect ratios considered if the atom number is low enough.

In Fig.~\ref{Fig:PDsmallN} we present a slice of the phase diagram in ($N,a_s$)-space. The phase boundary of the dysprosium results [Fig.~\ref{Fig:PDsmallN}(c)] corresponds to the magenta line on the phase diagram shown in Fig.~\ref{FigQFPD} (noting that it appears almost vertical in Fig.~\ref{FigQFPD} because of the $N$ scale used there).  In Figs.~\ref{Fig:PDsmallN}(a) and (b) we consider the peak density as a function of $a_s$ for condensates of various numbers in a prolate trap. For this trap geometry there is no phase transition at large $N$ (just a continuous cross over from the LDP to the HDP as $a_s$ reduces), but we observe that distinct LDP and HDP branches emerge for sufficiently low atom number. At the  aspect ratio considered here, the critical point  is at $N\approx1100$, $a_s\approx71\,a_0$ for $^{164}$Dy and $N\approx3400$, $a_s\approx44\,a_0$  for $^{168}$Er.

  \begin{figure}[htbp]
   \centering
  \includegraphics[width=3.3in]{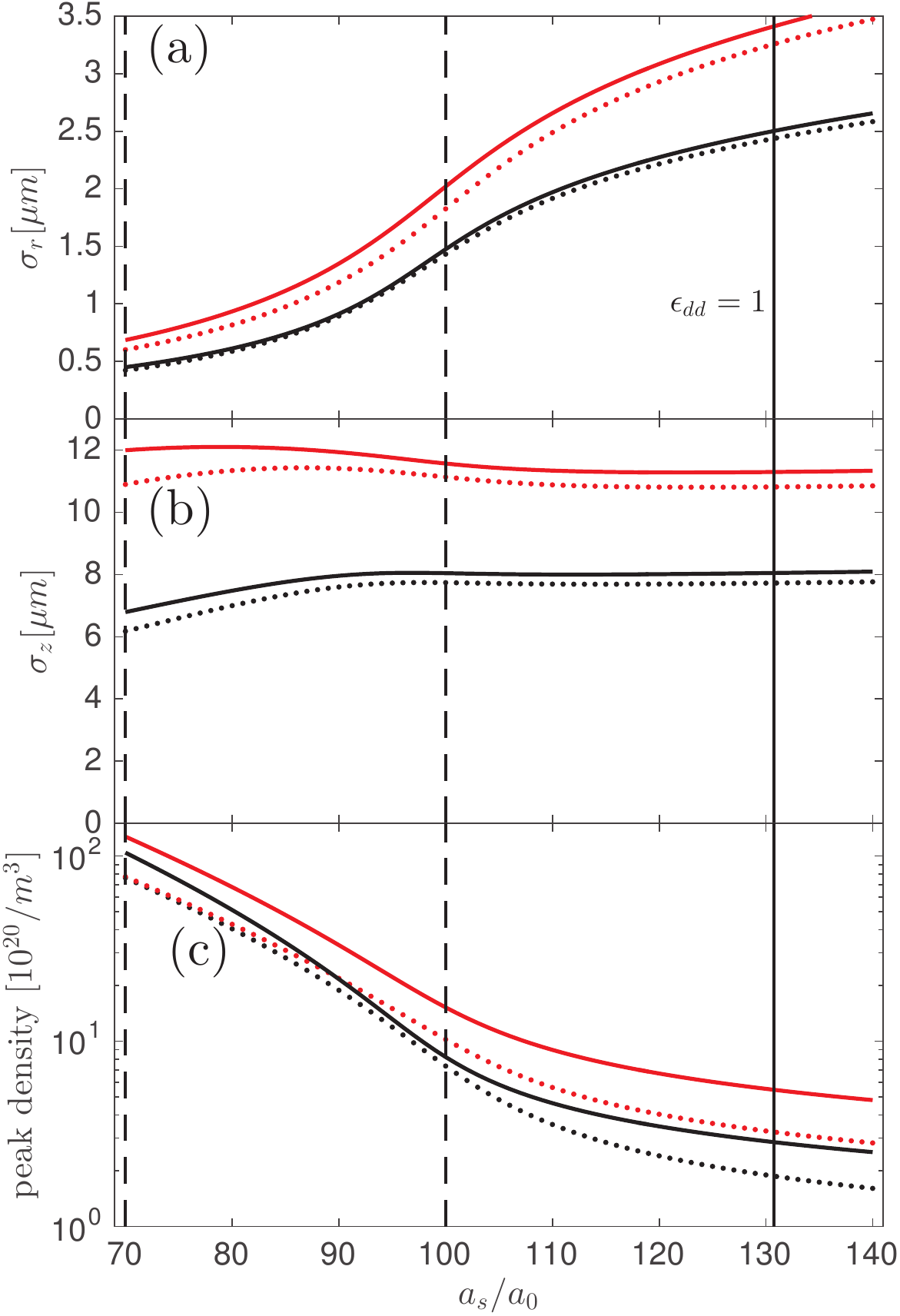}
  \vspace*{-0.05cm}
   \caption{ Properties of a single droplet of Dy atoms in a prolate trap as a function of $a_s$: (a) radial width; (b) axial width; (c) peak density. Variational solutions are given as lines while GPE results are shown as points. Black is for $N=10^4$ atoms and red (gray), for $N=5\times 10^4$. Vertical dashed lines represent the scattering lengths of the curves shown in Fig.~\ref{Fig:Dy_N}. Vertical solid line marks $\epsilon_{\mathrm{dd}}=1$. Trap: $\omega_\rho=2\pi\times 81.4$ s$^{-1}$, $\omega_z=2\pi\times 40.7$ s$^{-1}$.
   }
   \label{Fig:Dy_as}
\end{figure} 

\begin{figure}[htbp]
   \centering
  \includegraphics[width=3.3in]{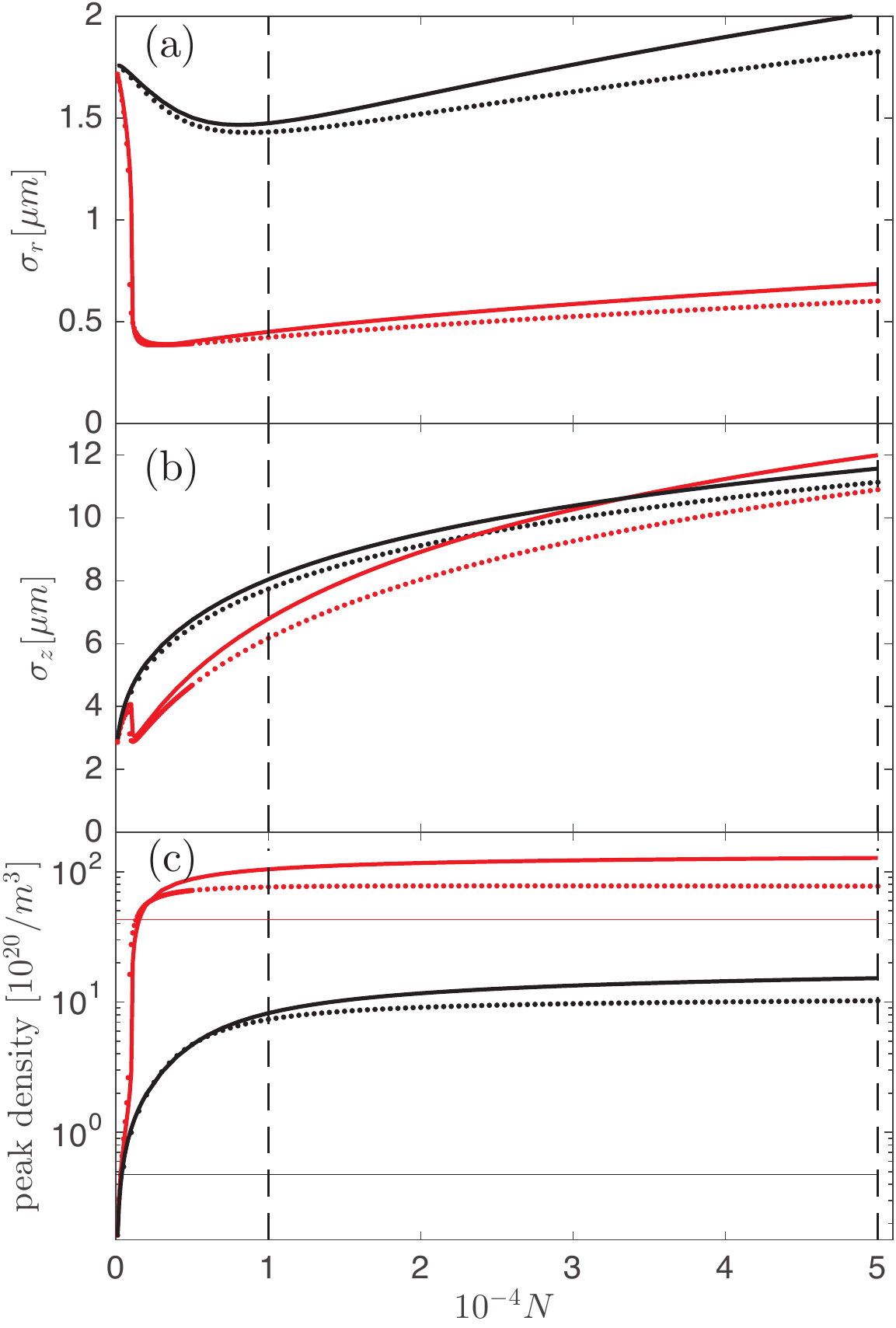} 
  \vspace*{-0.05cm}
   \caption{ Properties of a single droplet of Dy atoms in a prolate trap as a function of $N$: (a) radial width; (b) axial width; (c) peak density. Variational solutions are given as lines while GPE results are shown as points. Black is for $a_s/a_0=100$ and red (gray), for $a_s/a_0=70$. Thin horizontal lines show the analytic prediction of Eq.~(\ref{Eq:np_TF}). Vertical dashed lines represent the $N$ values of the curves in Fig.~\ref{Fig:Dy_as}. Trap: $\omega_\rho=2\pi\times 81.4$ s$^{-1}$, $\omega_z=2\pi\times 40.7$ s$^{-1}$. }
   \label{Fig:Dy_N}
\end{figure}

\subsection{Relationship to dynamics}\label{Secdynamics}
While the focus of  this paper is on the equilibrium states it is useful to relate our results to the dynamical simulations presented in Ref.~\cite{Wachtler2016a} that used the time-dependent generalized GPE (\ref{GPE}). The main scenario studied there was the quench used in experiments \cite{Kadau2016a} where the scattering length was reduced from an initial value $a_s=120\,a_0$  to the final value of $a_s=70\,a_0$ over a period of $0.5\,$ms. Such a quench takes the system across the phase boundary from the LDP to the HDP [this quench is indicated by the red dotted line in Fig.~\ref{Fig:PT_bigN}(a), also shown as the path labeled (1) in Fig.~\ref{FigQFPD}]. This quench does not produce the ground state we predict here (i.e.~a single droplet), instead approximately $10\,$ms after the quench a  droplet crystal consisting of up to 10 droplets forms. Each droplet contains approximately $10^3$ atoms. These observations are consistent with those made in the experiments. In simulations the number of droplets produced varies from run-to-run due to the  stochastic noise added to the LDP initial condition to mimic the effects of quantum and thermal fluctuations.  We conclude that the crystal forms as a metastable excited state because the system is unable to adiabatically cross the first order phase transition.
Indeed, as observed in Ref.~\cite{Wachtler2016a} (and similar to the findings in Ref.~\cite{Blakie2016a}) the system approximately conserves energy during the droplet formation (after the quench) and is unable to access the true ground state, which has a much lower energy. Instead, the crystal of droplets is consistent with this energy constraint, because each small (i.e.~containing a smaller number of atoms) droplet has a greater energy per particle than large droplets.
In Ref.~\cite{Wachtler2016a} it was reported that for $a_s=70\,a_0$ they were unable to solve for droplets with $N\lesssim900$ using an imaginary time method to solve for stationary states of the generalized GPE.  This is approximately the same value of $N$ that we find the phase boundary between the LDP and HDP phases, see Fig.~\ref{Fig:PDsmallN} for $a_s=70\,a_0$.

\subsection{Continuous droplet formation}
In experiments the properties of dipolar condensates are conveniently explored by changing the $s$-wave scattering length using Feshbach resonances. However, as discussed in Sec.~\ref{Secdynamics}, when a phase boundary is crossed the system will be unable to follow adiabatically and will end up in an excited state (e.g.~a droplet crystal). Aided by the phase diagrams we have developed here, we can explore paths in parameter space that go from the LDP to the HDP without crossing a phase boundary, i.e.~paths beyond the critical point [e.g.~the path labeled (2) in Fig.~\ref{FigQFPD}]. For example, with reference to Fig.~\ref{FigQFPD},  $a_s$ quenches with $N\gtrsim10^3$ and $\lambda\lesssim1$  do not cross a phase boundary, but continuously evolve from the LDP to the HDP. 

In this section we consider a particular case of a dipolar condensate in a trap of aspect ratio  $\lambda= 0.5$. 
Widths and peak densities are shown as a function of $a_s$ in Fig.~\ref{Fig:Dy_as} and as a function of $N$ in Fig.~\ref{Fig:Dy_N} for dysprosium condensates.  
The widths predicted by our variational model are shown as solid lines, and are in reasonable agreement with the full generalized GPE calculations, indicated by points, especially at small $N$.
As the scattering length is reduced, such that $\epsilon_{\mathrm{dd}}$ > 1, the droplet regime is smoothly obtained as indicated by the rapid radial shrinking of $\sigma_r$ [Fig.~\ref{Fig:Dy_as}(a)] and a corresponding increase of the peak density [Fig.~\ref{Fig:Dy_as}(c)]. Meanwhile, $\sigma_z$ remains relatively unaffected by changes of $a_s$.

Considering a different slice in parameter space, Fig.~\ref{Fig:Dy_N}(a) reveals that the radial width does not change monotonically with $N$, but instead exhibits a minimum at finite $N$ after which the droplet proceeds gradually to fatten with increasing $N$. While Fig.~\ref{Fig:Dy_as}(b) demonstrated that the axial width is relatively insensitive to changes of $a_s$, Fig.~\ref{Fig:Dy_N}(b) shows that $\sigma_z$ markedly increases with increasing $N$, with the exception of a small downward spike as the critical point is grazed at small $N$ for $a_s/a_0=70$.
Remarkably, Fig.~\ref{Fig:Dy_N}(c) demonstrates that the peak density saturates at a constant value as a function of $N$.
This saturation is consistent with the Thomas-Fermi prediction for the peak density in Ref.~\cite{Ferrier-Barbut2016a}
\begin{equation}
n_{\rm peak}^{\rm TF} = \frac{\pi}{a_s^3} \left( \frac{\epsilon_{\mathrm{dd}}f(\kappa)-1}{16(1+3\epsilon_{\mathrm{dd}}^2/2)} \right)^2 , \label{Eq:np_TF}
\end{equation}
where $\kappa = \sigma_r/\sigma_z$ is the droplet aspect ratio. We have drawn this prediction in Fig.~\ref{Fig:Dy_N} (c) as thin horizontal lines.
The agreement with our GPE and variational results is reasonable, deep in the droplet regime where $a_s/a_0=70$ ($\epsilon_{\mathrm{dd}} = 1.87$), but caution should be taken at larger scattering lengths as can be seen by the large deviation for the case of $a_s/a_0=100$ ($\epsilon_{\mathrm{dd}}=1.31$). The latter issue arises when $\epsilon_{\mathrm{dd}}f(\kappa)\approx 1$, resulting in a vanishing numerator of Eq.~(\ref{Eq:np_TF}) \footnote{For the case of $a_s/a_0=70$ we use $\kappa=1/20$, and for $a_s/a_0=100$ we use $\kappa=1/5$ as estimates of the droplet aspect ratios based on our results in Fig.~\ref{Fig:Dy_N} (a) and (b)}.
Peak density saturation should be favorable for future experiments, as it means that broad droplets with large atom numbers should be accessible without reaching densities where loss is problematic.
Importantly, large atom numbers in prolate traps should produce droplets with large \emph{in situ} widths, on the order of a few $\mu$m. Such widths are within reach of current \emph{in situ} imaging methods and experiments in this regime should furnish a fertile testbed to determine the precise details of the stabilization mechanism.

\section{Conclusions}
In this paper we have developed a phase diagram for the ground state of a dipolar condensate including the effect of quantum fluctuations. These fluctuations are treated within a local density approximation, found to be accurate in recent work that made comparisons to path-integral Monte Carlo simulations \cite{Saito2016a}.  This treatment, based on the leading order expression for the contribution of the quantum fluctuations to the energy, should be valid as long as the system is sufficiently dilute. The highest densities we encounter in our results are $n\sim10^{23}\,$m$^{-3}$, for which $na_{\mathrm{dd}}^3\sim0.03$  (using $a_{\mathrm{dd}}$ for $^{164}$Dy), thus the system is still reasonably dilute.
 
Our main results concentrate on the two types of stationary states that occur in this system:  a low-density condensate and a high-density droplet state that we identify as the LDP and HDP, respectively.
We have used a simple variational Gaussian ansatz that quantifies these states, and have validated the predictions of the variational solutions against full numerical solutions of the generalized GPE.
We find that the variational predictions are reasonably accurate, although we observe that in some regimes the full generalized GPE solutions for the HDP develop novel halo features not captured by the variational solution. We find that this halo arises from a local minimum in the effective potential, which should also confine excited thermal atoms, and thus may be observable in experiments.

 In general there is a first order phase transition between the LDP and HDP, although beyond the critical line is it possible to smoothly go between the phases. We have examined the phase diagram as a function of $a_s$, $N$ and $\lambda$, and explored the generic behavior of the phase transition for the cases of small and large $N$. 
Dynamical simulations have shown that the quench used in experiments crosses the phase boundary into the HDP non-adiabatically, and hence an excited state (a droplet crystal) is produced rather than the HDP ground state of a single droplet. Utilizing our phase diagram we propose using a similar quench in a dipolar condensate confined by a prolate trapping potential. This case lies beyond the critical line, and the LDP continuously evolves to the HDP. Alternatively, one could directly evaporatively cool the atoms at low scattering length in the prolate trap. Our results show that a single droplet forms with parameters suitable for direct observation in experiments. This regime may provide an opportunity to more carefully quantify the role of quantum fluctuations in the ground state. In addition to considering $^{164}$Dy  we have also presented results for $^{168}$Er to show that despite its smaller dipole moment this atom is well suited for exploring our predicted phase diagram in experiments.

\section{Acknowledgments}
We would like to thank F.~Dalfovo and S.~Stringari for useful discussions. RNB acknowledges support by the QUIC grant of the Horizon2020 FET program and by Provincia Autonoma di Trento. RMW acknowledges partial support from the National Science Foundation under Grant No.~PHYS-1516421. DB and PBB acknowledge the contribution of NZ eScience Infrastructure (NeSI) high-performance computing facilities, and support from the Marsden Fund of the Royal Society of New Zealand.

 \appendix
 \section{Quantum Fluctuations}
In this appendix we examine the calculation of the quantum fluctuation corrections motivating the choice of the parameter $\gamma_{\mathrm{QF}}$ we make in Eq.~(\ref{gammaV}). 
Other papers in the literature have made slightly different choices for evaluating this parameter (notably treating finite size effects), and for clarity we discuss and compare these other choices.
We begin by reviewing the calculation in the homogeneous case before turning to the extension of this result to a finite system.

\subsection{Homogeneous system}
We consider a uniform dipolar condensate system with Bogoliubov spectrum
\begin{align}
    \epsilon_\bk = \sqrt{\epsilon^0_k [\epsilon^0_k + 2ngf(\vartheta)] },
\end{align}
where $\epsilon^0_k=\hbar^2k^2/2m$, $f(\vartheta) = 1+\edd(3\cos^2\vartheta-1)$ and $\vartheta$ is the angle between the polarization of the dipoles and $\bk$. The corrections to the energy and chemical potential due to quantum fluctuations are given by \cite{Lima2012a}\footnote{The last term of Eq.~(29) of \cite{Lima2012a} is too high by a factor of $2$}
\begin{align}
    \Delta E &= \frac{V}2\int\frac{d\bk}{(2\pi)^{3}}  \left[\epsilon_\bk-\epsilon_k^{0}-ngf(\vartheta)+\frac{[ngf(\vartheta)]^{2}}{2\epsilon_k^{0}}\right],\label{e:DEfull}\\
    &= \frac{64Vg}{15} \sqrt{\frac{a_{s}^{3}}{\pi}}\mathcal{Q}_5(\edd) n^{5/2} = \frac{2V}5\gamma_{\mathrm{QF}} n^{5/2} ,\label{e:DEBog}\\
 \Delta \mu &= \frac{\partial \Delta E}{\partial N} =\gamma_{\mathrm{QF}} n^{3/2},\label{e:DmuBog}
 \end{align}
 where we have introduced the quantum fluctuation parameter
  \begin{equation}
\gamma_{\mathrm{QF}}\equiv   \frac{32g}{3} \sqrt{\frac{a_{s}^{3}}{\pi}}\mathcal{Q}_5(\edd),\label{gammaQ5}
  \end{equation}
 and $\mathcal{Q}_5(\edd)$ is given by
 \begin{align}
  \mathcal{Q}_5(\edd) 
 &= \frac{(3\edd)^{5/2}}{48}\Biggl[ (8 + 26y + 33y^2)\sqrt{1 + y} \notag\\
 &\hspace{2cm}+ 15y^3\ln\frac{1 + \sqrt{1 + y}}{\sqrt{y}} \Biggr],\label{e:Q5exact}\\
 &= 1+\frac32 \edd^2 + O(\edd^4),\label{e:Q5approx}
\end{align}
%
with $y=(1-\edd)/3\edd$. 

\subsection{Finite-size system}

\begin{figure}[htbp]
   \centering
  \includegraphics[width=3.3in]{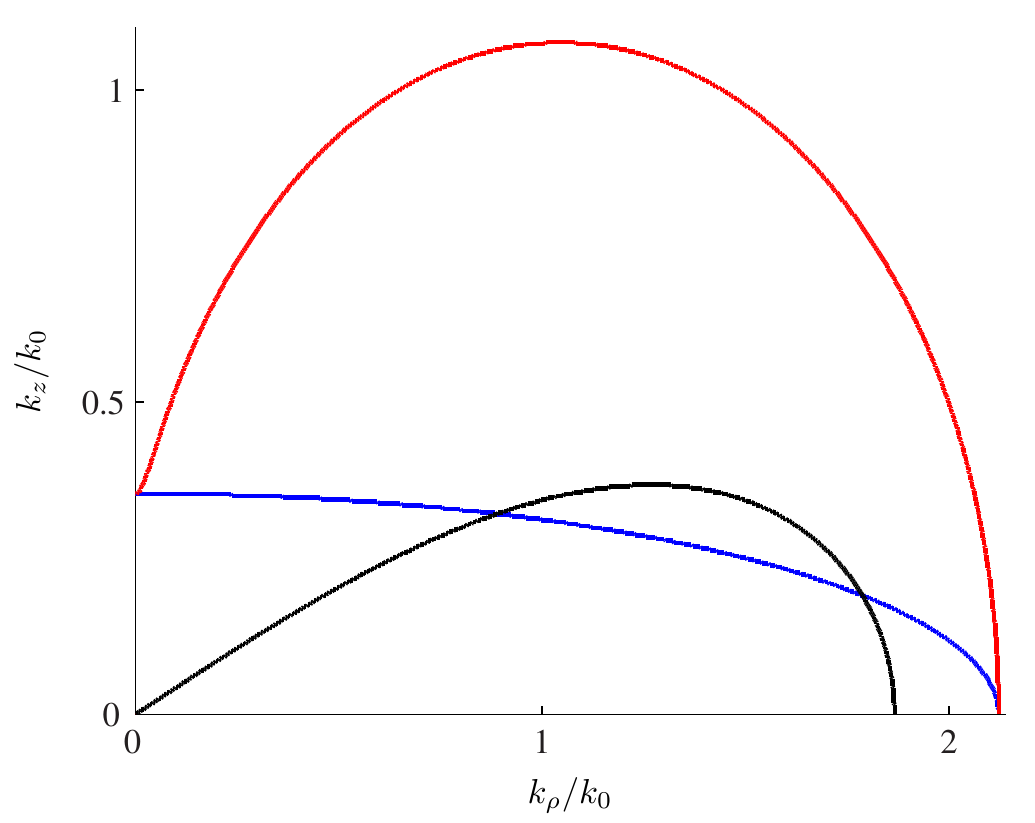} 
  \vspace*{0cm}
  \caption{(color online) Cutoff regions with $\{\sigma_\rho,\sigma_z\} = \{1/6,1\}\mu$m with $k_0$ from $n=1.5\times 10^{15}$cm$^{-3}$ and $a_s=70a_0$ \cite{Wachtler2016a} giving $\{k_{c,\rho},k_{z,\rho}\}=\{2.1,0.35\}k_0$. Lines indicate the $k_c$-boundaries (see text) used for \textit{Cutoff I} (blue/grey),    \textit{Cutoff II} (red/light grey), and \textit{Cutoff III} (with $\edd=1.87$, black).
 \label{f:kcutRegion} } 
\end{figure} 

In applying the homogeneous results to the experimental regime of micrometer scale droplets it is necessary to consider finite size effects. I.e., a system of extent $\{\sigma_\rho,\sigma_z\}$ can only support excitations of wavevectors exceeding the cutoff values $k_{c,i}=\pi/\sigma_i$, with $i=\{\rho,z\}$. We can account for this by restricting the $\mathbf{k}$-domain of the integral (\ref{e:DEfull}) to values greater than this.

\paragraph*{Cutoff I:} An obvious choice is to exclude an ellipsoidal region of long-wavelength modes, so that the integration is taken over the region 
\begin{equation}
(k_\rho/k_{c,\rho})^2+(k_z/k_{c,z})^2\ge 1,
\end{equation}
 (see Fig.~\ref{f:kcutRegion}), i.e.~an angular varying cutoff of the form 
\begin{equation}
k_c(\vartheta)=1/\sqrt{\sin^2\vartheta/k_{c,\rho}^2+\cos^2\vartheta/k_{c,z}^2}.
\end{equation} 
Performing this integration gives the parameter $\gamma_{\mathrm{QF}}$ according to Eq.~(\ref{gammaQ5}), but with $\mathcal{Q}_5$ replaced by $\mathcal{Q}'_5$ which can be calculated numerically as \footnote{This is consistent with Eq.~5 of \cite{Wachtler2016a}, except that their $-q_c(\theta)^2/2$ should be $-q_c(\theta)^2/10$.}
\begin{align}
    \mathcal{Q}'_5(\edd)  &= \frac1{64}\int_0^\pi \sin\vartheta d\vartheta [(8f(\vartheta)-3\bar{k}^2)(4f(\vartheta)+\bar{k}^2)^{3/2}\notag\\
    &\hspace{12mm}+3\bar{k}^5+10\bar{k}^3f(\vartheta)-30\bar{k}f(\vartheta)^2],\label{e:Q5p}
\end{align}
where $\bar{k} \equiv k_c(\vartheta)/k_0$ is the boundary of the integration region with $k_0=\sqrt{4\pi n a_s}$.

Two other cutoff choices have also appeared in the literature and we introduce these for the purposes of comparison.
\paragraph*{Cutoff II:} W{\"a}chtler \textit{et al.}~\cite{Wachtler2016a} used a cutoff of the form
\begin{equation}
 k_c(\vartheta) = \sqrt{k^2_{c,\rho}\sin^2\vartheta+ k^2_{c,z}\cos^2\vartheta},
\end{equation}
 (see Fig.~\ref{f:kcutRegion}), which can similarly be used to obtain $\gamma_{\mathrm{QF}}$ by numerically evaluating $\mathcal{Q}'_5$ as in Eq.~\eqref{e:Q5p}.
\paragraph*{Cutoff III:} For $\epsilon_{\mathrm{dd}}>1$ the integrand (\ref{e:DEfull}) is not real everywhere as some modes have gone soft and developed imaginary energies. Saito~\cite{Saito2016a} proposed restricting the domain of integration to regions where the integrand is real, i.e.~
\begin{equation}
k_c(\vartheta) =2\sqrt{\max[0,-gnmf(\vartheta)]}/\hbar,
\end{equation}
(see Fig.~\ref{f:kcutRegion}), which can be used to numerically evaluate $\mathcal{Q}'_5$ again using Eq.~\eqref{e:Q5p}.

In Fig.~\ref{f:RQ5d} we compare the results for $\mathcal{Q}'_5$ using the various cutoff procedures to the homogeneous result $\mathcal{Q}_5$, and the analytic approximation (\ref{e:Q5approx}) that we use in this paper. These results are for the same size droplets considered in Ref.~\cite{Wachtler2016a}.
These results clearly show that the cutoff procedure has little effect on the value of $\mathcal{Q}'_5$ over the homogeneous result, and clearly the simple analytic approximation (\ref{e:Q5exact}) is justified [which is the basis of our choice for $\gamma_{\mathrm{QF}}$ in this work, see Eq.~({\ref{gammaV}), also see \cite{Ferrier-Barbut2016a}].
Consistent with other treatments (see \cite{Ferrier-Barbut2016a,Saito2016a,Wachtler2016a}) we neglect any (typically small) imaginary part in $\mathcal{Q}_5$, arising from the soft modes for $\epsilon_{\mathrm{dd}}>1$.

\begin{figure}[htbp]
   \centering
  \includegraphics[width=3.3in]{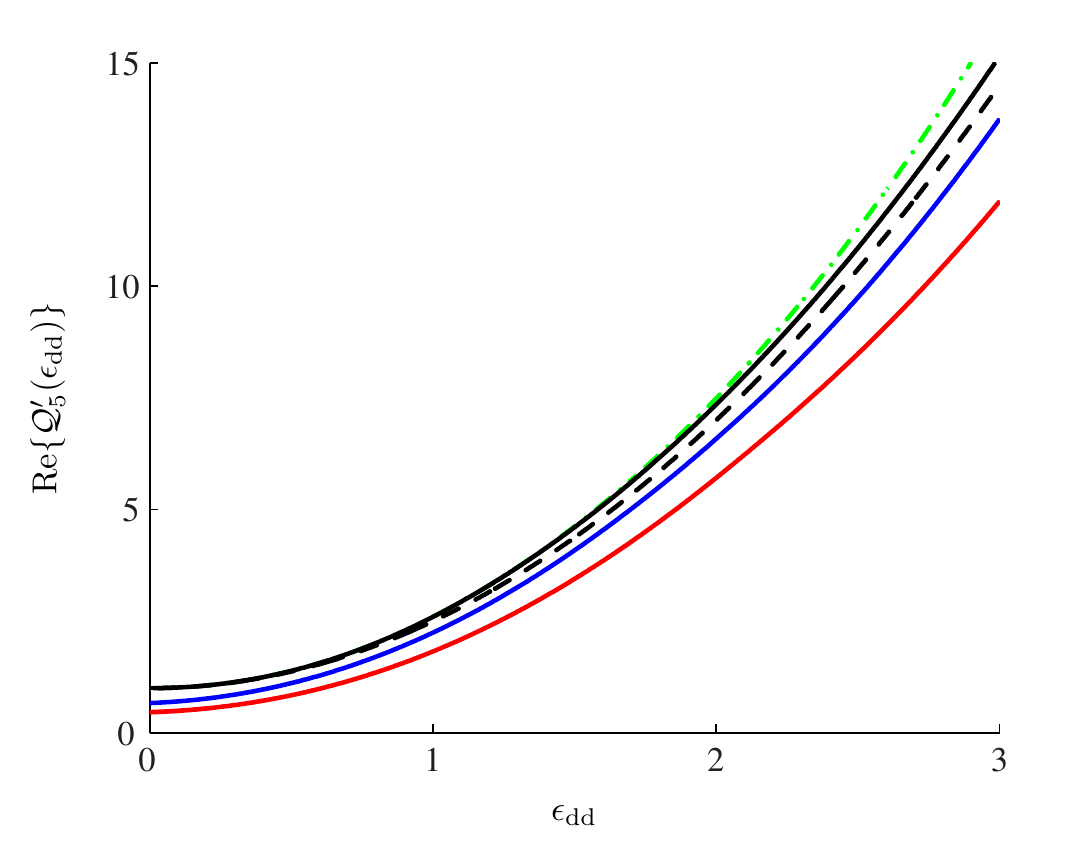} 
  \vspace*{0cm}
  \caption{(color online)  Results for the real part of $\mathcal{Q}_5^\prime(\edd)$ for the cutoff options from 
  Fig.~\ref{f:kcutRegion} (same colors). Also shown is the result with no cutoff (green dash-dotted) and the approximate result $\mathcal{Q}_5(\edd) = 1+\frac32 \edd^2$ (black dashed). At $\edd=1.87$, $\mathrm{Im}\{\mathcal{Q}_5(\edd)\}\approx 0.137$ and $\mathrm{Im}\{\mathcal{Q}^\prime_5(\edd)\}\approx 0.013$ for \textit{Cutoff I}.\label{f:RQ5d} }
\end{figure}


\begin{thebibliography}{53}%
\makeatletter
\providecommand \@ifxundefined [1]{%
 \@ifx{#1\undefined}
}%
\providecommand \@ifnum [1]{%
 \ifnum #1\expandafter \@firstoftwo
 \else \expandafter \@secondoftwo
 \fi
}%
\providecommand \@ifx [1]{%
 \ifx #1\expandafter \@firstoftwo
 \else \expandafter \@secondoftwo
 \fi
}%
\providecommand \natexlab [1]{#1}%
\providecommand \enquote  [1]{``#1''}%
\providecommand \bibnamefont  [1]{#1}%
\providecommand \bibfnamefont [1]{#1}%
\providecommand \citenamefont [1]{#1}%
\providecommand \href@noop [0]{\@secondoftwo}%
\providecommand \href [0]{\begingroup \@sanitize@url \@href}%
\providecommand \@href[1]{\@@startlink{#1}\@@href}%
\providecommand \@@href[1]{\endgroup#1\@@endlink}%
\providecommand \@sanitize@url [0]{\catcode `\\12\catcode `\$12\catcode
  `\&12\catcode `\#12\catcode `\^12\catcode `\_12\catcode `\%12\relax}%
\providecommand \@@startlink[1]{}%
\providecommand \@@endlink[0]{}%
\providecommand \url  [0]{\begingroup\@sanitize@url \@url }%
\providecommand \@url [1]{\endgroup\@href {#1}{\urlprefix }}%
\providecommand \urlprefix  [0]{URL }%
\providecommand \Eprint [0]{\href }%
\providecommand \doibase [0]{http://dx.doi.org/}%
\providecommand \selectlanguage [0]{\@gobble}%
\providecommand \bibinfo  [0]{\@secondoftwo}%
\providecommand \bibfield  [0]{\@secondoftwo}%
\providecommand \translation [1]{[#1]}%
\providecommand \BibitemOpen [0]{}%
\providecommand \bibitemStop [0]{}%
\providecommand \bibitemNoStop [0]{.\EOS\space}%
\providecommand \EOS [0]{\spacefactor3000\relax}%
\providecommand \BibitemShut  [1]{\csname bibitem#1\endcsname}%
\let\auto@bib@innerbib\@empty
\bibitem [{\citenamefont {Griesmaier}\ \emph {et~al.}(2005)\citenamefont
  {Griesmaier}, \citenamefont {Werner}, \citenamefont {Hensler}, \citenamefont
  {Stuhler},\ and\ \citenamefont {Pfau}}]{Griesmaier2005a}%
  \BibitemOpen
  \bibfield  {author} {\bibinfo {author} {\bibfnamefont {A.}~\bibnamefont
  {Griesmaier}}, \bibinfo {author} {\bibfnamefont {J.}~\bibnamefont {Werner}},
  \bibinfo {author} {\bibfnamefont {S.}~\bibnamefont {Hensler}}, \bibinfo
  {author} {\bibfnamefont {J.}~\bibnamefont {Stuhler}}, \ and\ \bibinfo
  {author} {\bibfnamefont {T.}~\bibnamefont {Pfau}},\ }\href {\doibase
  10.1103/PhysRevLett.94.160401} {\bibfield  {journal} {\bibinfo  {journal}
  {Phys. Rev. Lett.}\ }\textbf {\bibinfo {volume} {94}},\ \bibinfo {pages}
  {160401} (\bibinfo {year} {2005})}\BibitemShut {NoStop}%
\bibitem [{\citenamefont {Lu}\ \emph {et~al.}(2011)\citenamefont {Lu},
  \citenamefont {Burdick}, \citenamefont {Youn},\ and\ \citenamefont
  {Lev}}]{Mingwu2011a}%
  \BibitemOpen
  \bibfield  {author} {\bibinfo {author} {\bibfnamefont {M.}~\bibnamefont
  {Lu}}, \bibinfo {author} {\bibfnamefont {N.~Q.}\ \bibnamefont {Burdick}},
  \bibinfo {author} {\bibfnamefont {S.~H.}\ \bibnamefont {Youn}}, \ and\
  \bibinfo {author} {\bibfnamefont {B.~L.}\ \bibnamefont {Lev}},\ }\href
  {\doibase 10.1103/PhysRevLett.107.190401} {\bibfield  {journal} {\bibinfo
  {journal} {Phys. Rev. Lett.}\ }\textbf {\bibinfo {volume} {107}},\ \bibinfo
  {pages} {190401} (\bibinfo {year} {2011})}\BibitemShut {NoStop}%
\bibitem [{\citenamefont {Aikawa}\ \emph {et~al.}(2012)\citenamefont {Aikawa},
  \citenamefont {Frisch}, \citenamefont {Mark}, \citenamefont {Baier},
  \citenamefont {Rietzler}, \citenamefont {Grimm},\ and\ \citenamefont
  {Ferlaino}}]{Aikawa2012a}%
  \BibitemOpen
  \bibfield  {author} {\bibinfo {author} {\bibfnamefont {K.}~\bibnamefont
  {Aikawa}}, \bibinfo {author} {\bibfnamefont {A.}~\bibnamefont {Frisch}},
  \bibinfo {author} {\bibfnamefont {M.}~\bibnamefont {Mark}}, \bibinfo {author}
  {\bibfnamefont {S.}~\bibnamefont {Baier}}, \bibinfo {author} {\bibfnamefont
  {A.}~\bibnamefont {Rietzler}}, \bibinfo {author} {\bibfnamefont
  {R.}~\bibnamefont {Grimm}}, \ and\ \bibinfo {author} {\bibfnamefont
  {F.}~\bibnamefont {Ferlaino}},\ }\href {\doibase
  10.1103/PhysRevLett.108.210401} {\bibfield  {journal} {\bibinfo  {journal}
  {Phys. Rev. Lett.}\ }\textbf {\bibinfo {volume} {108}},\ \bibinfo {pages}
  {210401} (\bibinfo {year} {2012})}\BibitemShut {NoStop}%
\bibitem [{\citenamefont {Lahaye}\ \emph {et~al.}(2009)\citenamefont {Lahaye},
  \citenamefont {Menotti}, \citenamefont {Santos}, \citenamefont {Lewenstein},\
  and\ \citenamefont {Pfau}}]{Lahaye_RepProgPhys_2009}%
  \BibitemOpen
  \bibfield  {author} {\bibinfo {author} {\bibfnamefont {T.}~\bibnamefont
  {Lahaye}}, \bibinfo {author} {\bibfnamefont {C.}~\bibnamefont {Menotti}},
  \bibinfo {author} {\bibfnamefont {L.}~\bibnamefont {Santos}}, \bibinfo
  {author} {\bibfnamefont {M.}~\bibnamefont {Lewenstein}}, \ and\ \bibinfo
  {author} {\bibfnamefont {T.}~\bibnamefont {Pfau}},\ }\href
  {http://stacks.iop.org/0034-4885/72/126401} {\bibfield  {journal} {\bibinfo
  {journal} {Rep. Prog. Phys.}\ }\textbf {\bibinfo {volume} {72}},\ \bibinfo
  {pages} {126401} (\bibinfo {year} {2009})}\BibitemShut {NoStop}%
\bibitem [{\citenamefont {Santos}\ \emph {et~al.}(2003)\citenamefont {Santos},
  \citenamefont {Shlyapnikov},\ and\ \citenamefont {Lewenstein}}]{Santos2003a}%
  \BibitemOpen
  \bibfield  {author} {\bibinfo {author} {\bibfnamefont {L.}~\bibnamefont
  {Santos}}, \bibinfo {author} {\bibfnamefont {G.~V.}\ \bibnamefont
  {Shlyapnikov}}, \ and\ \bibinfo {author} {\bibfnamefont {M.}~\bibnamefont
  {Lewenstein}},\ }\href {\doibase 10.1103/PhysRevLett.90.250403} {\bibfield
  {journal} {\bibinfo  {journal} {Phys. Rev. Lett.}\ }\textbf {\bibinfo
  {volume} {90}},\ \bibinfo {pages} {250403} (\bibinfo {year}
  {2003})}\BibitemShut {NoStop}%
\bibitem [{\citenamefont {Ronen}\ \emph
  {et~al.}(2006{\natexlab{a}})\citenamefont {Ronen}, \citenamefont
  {Bortolotti}, \citenamefont {Blume},\ and\ \citenamefont
  {Bohn}}]{Ronen2006b}%
  \BibitemOpen
  \bibfield  {author} {\bibinfo {author} {\bibfnamefont {S.}~\bibnamefont
  {Ronen}}, \bibinfo {author} {\bibfnamefont {D.~C.~E.}\ \bibnamefont
  {Bortolotti}}, \bibinfo {author} {\bibfnamefont {D.}~\bibnamefont {Blume}}, \
  and\ \bibinfo {author} {\bibfnamefont {J.~L.}\ \bibnamefont {Bohn}},\ }\href
  {\doibase 10.1103/PhysRevA.74.033611} {\bibfield  {journal} {\bibinfo
  {journal} {Phys. Rev. A}\ }\textbf {\bibinfo {volume} {74}},\ \bibinfo {eid}
  {033611} (\bibinfo {year} {2006}{\natexlab{a}})}\BibitemShut {NoStop}%
\bibitem [{\citenamefont {Ronen}\ \emph
  {et~al.}(2006{\natexlab{b}})\citenamefont {Ronen}, \citenamefont
  {Bortolotti},\ and\ \citenamefont {Bohn}}]{Ronen2006a}%
  \BibitemOpen
  \bibfield  {author} {\bibinfo {author} {\bibfnamefont {S.}~\bibnamefont
  {Ronen}}, \bibinfo {author} {\bibfnamefont {D.~C.~E.}\ \bibnamefont
  {Bortolotti}}, \ and\ \bibinfo {author} {\bibfnamefont {J.~L.}\ \bibnamefont
  {Bohn}},\ }\href {\doibase 10.1103/PhysRevA.74.013623} {\bibfield  {journal}
  {\bibinfo  {journal} {Phys. Rev. A}\ }\textbf {\bibinfo {volume} {74}},\
  \bibinfo {pages} {013623} (\bibinfo {year} {2006}{\natexlab{b}})}\BibitemShut
  {NoStop}%
\bibitem [{\citenamefont {Koch}\ \emph {et~al.}(2008)\citenamefont {Koch},
  \citenamefont {Lahaye}, \citenamefont {Metz}, \citenamefont {Frohlich},
  \citenamefont {Griesmaier},\ and\ \citenamefont {Pfau}}]{Koch2008a}%
  \BibitemOpen
  \bibfield  {author} {\bibinfo {author} {\bibfnamefont {T.}~\bibnamefont
  {Koch}}, \bibinfo {author} {\bibfnamefont {T.}~\bibnamefont {Lahaye}},
  \bibinfo {author} {\bibfnamefont {J.}~\bibnamefont {Metz}}, \bibinfo {author}
  {\bibfnamefont {B.}~\bibnamefont {Frohlich}}, \bibinfo {author}
  {\bibfnamefont {A.}~\bibnamefont {Griesmaier}}, \ and\ \bibinfo {author}
  {\bibfnamefont {T.}~\bibnamefont {Pfau}},\ }\href {\doibase 10.1038/nphys887}
  {\bibfield  {journal} {\bibinfo  {journal} {Nat. Phys.}\ }\textbf {\bibinfo
  {volume} {4}},\ \bibinfo {pages} {218} (\bibinfo {year} {2008})}\BibitemShut
  {NoStop}%
\bibitem [{\citenamefont {Ticknor}\ \emph {et~al.}(2008)\citenamefont
  {Ticknor}, \citenamefont {Parker}, \citenamefont {Melatos}, \citenamefont
  {Cornish}, \citenamefont {O'Dell},\ and\ \citenamefont
  {Martin}}]{Ticknor2008a}%
  \BibitemOpen
  \bibfield  {author} {\bibinfo {author} {\bibfnamefont {C.}~\bibnamefont
  {Ticknor}}, \bibinfo {author} {\bibfnamefont {N.~G.}\ \bibnamefont {Parker}},
  \bibinfo {author} {\bibfnamefont {A.}~\bibnamefont {Melatos}}, \bibinfo
  {author} {\bibfnamefont {S.~L.}\ \bibnamefont {Cornish}}, \bibinfo {author}
  {\bibfnamefont {D.~H.~J.}\ \bibnamefont {O'Dell}}, \ and\ \bibinfo {author}
  {\bibfnamefont {A.~M.}\ \bibnamefont {Martin}},\ }\href {\doibase
  10.1103/PhysRevA.78.061607} {\bibfield  {journal} {\bibinfo  {journal} {Phys.
  Rev. A}\ }\textbf {\bibinfo {volume} {78}},\ \bibinfo {eid} {061607}
  (\bibinfo {year} {2008})}\BibitemShut {NoStop}%
\bibitem [{\citenamefont {Lahaye}\ \emph {et~al.}(2008)\citenamefont {Lahaye},
  \citenamefont {Metz}, \citenamefont {Fr\"{o}hlich}, \citenamefont {Koch},
  \citenamefont {Meister}, \citenamefont {Griesmaier}, \citenamefont {Pfau},
  \citenamefont {Saito}, \citenamefont {Kawaguchi},\ and\ \citenamefont
  {Ueda}}]{Lahaye2009a}%
  \BibitemOpen
  \bibfield  {author} {\bibinfo {author} {\bibfnamefont {T.}~\bibnamefont
  {Lahaye}}, \bibinfo {author} {\bibfnamefont {J.}~\bibnamefont {Metz}},
  \bibinfo {author} {\bibfnamefont {B.}~\bibnamefont {Fr\"{o}hlich}}, \bibinfo
  {author} {\bibfnamefont {T.}~\bibnamefont {Koch}}, \bibinfo {author}
  {\bibfnamefont {M.}~\bibnamefont {Meister}}, \bibinfo {author} {\bibfnamefont
  {A.}~\bibnamefont {Griesmaier}}, \bibinfo {author} {\bibfnamefont
  {T.}~\bibnamefont {Pfau}}, \bibinfo {author} {\bibfnamefont {H.}~\bibnamefont
  {Saito}}, \bibinfo {author} {\bibfnamefont {Y.}~\bibnamefont {Kawaguchi}}, \
  and\ \bibinfo {author} {\bibfnamefont {M.}~\bibnamefont {Ueda}},\ }\href
  {\doibase 10.1103/PhysRevLett.101.080401} {\bibfield  {journal} {\bibinfo
  {journal} {Phys. Rev. Lett.}\ }\textbf {\bibinfo {volume} {101}},\ \bibinfo
  {eid} {080401} (\bibinfo {year} {2008})}\BibitemShut {NoStop}%
\bibitem [{\citenamefont {Parker}\ \emph {et~al.}(2009)\citenamefont {Parker},
  \citenamefont {Ticknor}, \citenamefont {Martin},\ and\ \citenamefont
  {O'Dell}}]{Parker2009a}%
  \BibitemOpen
  \bibfield  {author} {\bibinfo {author} {\bibfnamefont {N.~G.}\ \bibnamefont
  {Parker}}, \bibinfo {author} {\bibfnamefont {C.}~\bibnamefont {Ticknor}},
  \bibinfo {author} {\bibfnamefont {A.~M.}\ \bibnamefont {Martin}}, \ and\
  \bibinfo {author} {\bibfnamefont {D.~H.~J.}\ \bibnamefont {O'Dell}},\ }\href
  {\doibase 10.1103/PhysRevA.79.013617} {\bibfield  {journal} {\bibinfo
  {journal} {Phys. Rev. A}\ }\textbf {\bibinfo {volume} {79}},\ \bibinfo {eid}
  {013617} (\bibinfo {year} {2009})}\BibitemShut {NoStop}%
\bibitem [{\citenamefont {Wilson}\ \emph {et~al.}(2009)\citenamefont {Wilson},
  \citenamefont {Ronen},\ and\ \citenamefont {Bohn}}]{Wilson2009a}%
  \BibitemOpen
  \bibfield  {author} {\bibinfo {author} {\bibfnamefont {R.~M.}\ \bibnamefont
  {Wilson}}, \bibinfo {author} {\bibfnamefont {S.}~\bibnamefont {Ronen}}, \
  and\ \bibinfo {author} {\bibfnamefont {J.~L.}\ \bibnamefont {Bohn}},\ }\href
  {\doibase 10.1103/PhysRevA.80.023614} {\bibfield  {journal} {\bibinfo
  {journal} {Phys. Rev. A}\ }\textbf {\bibinfo {volume} {80}},\ \bibinfo
  {pages} {023614} (\bibinfo {year} {2009})}\BibitemShut {NoStop}%
\bibitem [{\citenamefont {Lu}\ \emph {et~al.}(2010)\citenamefont {Lu},
  \citenamefont {Lu}, \citenamefont {Zhang}, \citenamefont {Qiu}, \citenamefont
  {Pu},\ and\ \citenamefont {Yi}}]{Lu2010a}%
  \BibitemOpen
  \bibfield  {author} {\bibinfo {author} {\bibfnamefont {H.-Y.}\ \bibnamefont
  {Lu}}, \bibinfo {author} {\bibfnamefont {H.}~\bibnamefont {Lu}}, \bibinfo
  {author} {\bibfnamefont {J.-N.}\ \bibnamefont {Zhang}}, \bibinfo {author}
  {\bibfnamefont {R.-Z.}\ \bibnamefont {Qiu}}, \bibinfo {author} {\bibfnamefont
  {H.}~\bibnamefont {Pu}}, \ and\ \bibinfo {author} {\bibfnamefont
  {S.}~\bibnamefont {Yi}},\ }\href {\doibase 10.1103/PhysRevA.82.023622}
  {\bibfield  {journal} {\bibinfo  {journal} {Phys. Rev. A}\ }\textbf {\bibinfo
  {volume} {82}},\ \bibinfo {pages} {023622} (\bibinfo {year}
  {2010})}\BibitemShut {NoStop}%
\bibitem [{\citenamefont {M\"uller}\ \emph {et~al.}(2011)\citenamefont
  {M\"uller}, \citenamefont {Billy}, \citenamefont {Henn}, \citenamefont
  {Kadau}, \citenamefont {Griesmaier}, \citenamefont {Jona-Lasinio},
  \citenamefont {Santos},\ and\ \citenamefont {Pfau}}]{Muller2011a}%
  \BibitemOpen
  \bibfield  {author} {\bibinfo {author} {\bibfnamefont {S.}~\bibnamefont
  {M\"uller}}, \bibinfo {author} {\bibfnamefont {J.}~\bibnamefont {Billy}},
  \bibinfo {author} {\bibfnamefont {E.~A.~L.}\ \bibnamefont {Henn}}, \bibinfo
  {author} {\bibfnamefont {H.}~\bibnamefont {Kadau}}, \bibinfo {author}
  {\bibfnamefont {A.}~\bibnamefont {Griesmaier}}, \bibinfo {author}
  {\bibfnamefont {M.}~\bibnamefont {Jona-Lasinio}}, \bibinfo {author}
  {\bibfnamefont {L.}~\bibnamefont {Santos}}, \ and\ \bibinfo {author}
  {\bibfnamefont {T.}~\bibnamefont {Pfau}},\ }\href {\doibase
  10.1103/PhysRevA.84.053601} {\bibfield  {journal} {\bibinfo  {journal} {Phys.
  Rev. A}\ }\textbf {\bibinfo {volume} {84}},\ \bibinfo {pages} {053601}
  (\bibinfo {year} {2011})}\BibitemShut {NoStop}%
\bibitem [{\citenamefont {Bisset}\ \emph {et~al.}(2012)\citenamefont {Bisset},
  \citenamefont {Baillie},\ and\ \citenamefont {Blakie}}]{Bisset2012}%
  \BibitemOpen
  \bibfield  {author} {\bibinfo {author} {\bibfnamefont {R.~N.}\ \bibnamefont
  {Bisset}}, \bibinfo {author} {\bibfnamefont {D.}~\bibnamefont {Baillie}}, \
  and\ \bibinfo {author} {\bibfnamefont {P.~B.}\ \bibnamefont {Blakie}},\
  }\href {\doibase 10.1103/PhysRevA.86.033609} {\bibfield  {journal} {\bibinfo
  {journal} {Phys. Rev. A}\ }\textbf {\bibinfo {volume} {86}},\ \bibinfo
  {pages} {033609} (\bibinfo {year} {2012})}\BibitemShut {NoStop}%
\bibitem [{\citenamefont {Corson}\ \emph
  {et~al.}(2013{\natexlab{a}})\citenamefont {Corson}, \citenamefont {Wilson},\
  and\ \citenamefont {Bohn}}]{Corson2013a}%
  \BibitemOpen
  \bibfield  {author} {\bibinfo {author} {\bibfnamefont {J.~P.}\ \bibnamefont
  {Corson}}, \bibinfo {author} {\bibfnamefont {R.~M.}\ \bibnamefont {Wilson}},
  \ and\ \bibinfo {author} {\bibfnamefont {J.~L.}\ \bibnamefont {Bohn}},\
  }\href {\doibase 10.1103/PhysRevA.87.051605} {\bibfield  {journal} {\bibinfo
  {journal} {Phys. Rev. A}\ }\textbf {\bibinfo {volume} {87}},\ \bibinfo
  {pages} {051605} (\bibinfo {year} {2013}{\natexlab{a}})}\BibitemShut
  {NoStop}%
\bibitem [{\citenamefont {Corson}\ \emph
  {et~al.}(2013{\natexlab{b}})\citenamefont {Corson}, \citenamefont {Wilson},\
  and\ \citenamefont {Bohn}}]{Corson2013b}%
  \BibitemOpen
  \bibfield  {author} {\bibinfo {author} {\bibfnamefont {J.~P.}\ \bibnamefont
  {Corson}}, \bibinfo {author} {\bibfnamefont {R.~M.}\ \bibnamefont {Wilson}},
  \ and\ \bibinfo {author} {\bibfnamefont {J.~L.}\ \bibnamefont {Bohn}},\
  }\href {\doibase 10.1103/PhysRevA.88.013614} {\bibfield  {journal} {\bibinfo
  {journal} {Phys. Rev. A}\ }\textbf {\bibinfo {volume} {88}},\ \bibinfo
  {pages} {013614} (\bibinfo {year} {2013}{\natexlab{b}})}\BibitemShut
  {NoStop}%
\bibitem [{\citenamefont {Linscott}\ and\ \citenamefont
  {Blakie}(2014)}]{Linscott2014a}%
  \BibitemOpen
  \bibfield  {author} {\bibinfo {author} {\bibfnamefont {E.~B.}\ \bibnamefont
  {Linscott}}\ and\ \bibinfo {author} {\bibfnamefont {P.~B.}\ \bibnamefont
  {Blakie}},\ }\href {\doibase 10.1103/PhysRevA.90.053605} {\bibfield
  {journal} {\bibinfo  {journal} {Phys. Rev. A}\ }\textbf {\bibinfo {volume}
  {90}},\ \bibinfo {pages} {053605} (\bibinfo {year} {2014})}\BibitemShut
  {NoStop}%
\bibitem [{\citenamefont {Kadau}\ \emph {et~al.}(2016)\citenamefont {Kadau},
  \citenamefont {Schmitt}, \citenamefont {Wenzel}, \citenamefont {Wink},
  \citenamefont {Maier}, \citenamefont {Ferrier-Barbut},\ and\ \citenamefont
  {Pfau}}]{Kadau2016a}%
  \BibitemOpen
  \bibfield  {author} {\bibinfo {author} {\bibfnamefont {H.}~\bibnamefont
  {Kadau}}, \bibinfo {author} {\bibfnamefont {M.}~\bibnamefont {Schmitt}},
  \bibinfo {author} {\bibfnamefont {M.}~\bibnamefont {Wenzel}}, \bibinfo
  {author} {\bibfnamefont {C.}~\bibnamefont {Wink}}, \bibinfo {author}
  {\bibfnamefont {T.}~\bibnamefont {Maier}}, \bibinfo {author} {\bibfnamefont
  {I.}~\bibnamefont {Ferrier-Barbut}}, \ and\ \bibinfo {author} {\bibfnamefont
  {T.}~\bibnamefont {Pfau}},\ }\href {http://dx.doi.org/10.1038/nature16485}
  {\bibfield  {journal} {\bibinfo  {journal} {Nature}\ }\textbf {\bibinfo
  {volume} {530}},\ \bibinfo {pages} {194} (\bibinfo {year}
  {2016})}\BibitemShut {NoStop}%
\bibitem [{\citenamefont {Xi}\ and\ \citenamefont {Saito}(2016)}]{Xi2016a}%
  \BibitemOpen
  \bibfield  {author} {\bibinfo {author} {\bibfnamefont {K.-T.}\ \bibnamefont
  {Xi}}\ and\ \bibinfo {author} {\bibfnamefont {H.}~\bibnamefont {Saito}},\
  }\href {\doibase 10.1103/PhysRevA.93.011604} {\bibfield  {journal} {\bibinfo
  {journal} {Phys. Rev. A}\ }\textbf {\bibinfo {volume} {93}},\ \bibinfo
  {pages} {011604} (\bibinfo {year} {2016})}\BibitemShut {NoStop}%
\bibitem [{\citenamefont {Bisset}\ and\ \citenamefont
  {Blakie}(2015)}]{Bisset2015a}%
  \BibitemOpen
  \bibfield  {author} {\bibinfo {author} {\bibfnamefont {R.~N.}\ \bibnamefont
  {Bisset}}\ and\ \bibinfo {author} {\bibfnamefont {P.~B.}\ \bibnamefont
  {Blakie}},\ }\href {\doibase 10.1103/PhysRevA.92.061603} {\bibfield
  {journal} {\bibinfo  {journal} {Phys. Rev. A}\ }\textbf {\bibinfo {volume}
  {92}},\ \bibinfo {pages} {061603} (\bibinfo {year} {2015})}\BibitemShut
  {NoStop}%
\bibitem [{\citenamefont {Blakie}(2016)}]{Blakie2016a}%
  \BibitemOpen
  \bibfield  {author} {\bibinfo {author} {\bibfnamefont {P.~B.}\ \bibnamefont
  {Blakie}},\ }\href {\doibase 10.1103/PhysRevA.93.033644} {\bibfield
  {journal} {\bibinfo  {journal} {Phys. Rev. A}\ }\textbf {\bibinfo {volume}
  {93}},\ \bibinfo {pages} {033644} (\bibinfo {year} {2016})}\BibitemShut
  {NoStop}%
\bibitem [{\citenamefont {Petrov}(2014)}]{Petrov2014a}%
  \BibitemOpen
  \bibfield  {author} {\bibinfo {author} {\bibfnamefont {D.~S.}\ \bibnamefont
  {Petrov}},\ }\href {\doibase 10.1103/PhysRevLett.112.103201} {\bibfield
  {journal} {\bibinfo  {journal} {Phys. Rev. Lett.}\ }\textbf {\bibinfo
  {volume} {112}},\ \bibinfo {pages} {103201} (\bibinfo {year}
  {2014})}\BibitemShut {NoStop}%
\bibitem [{\citenamefont {Lu}\ \emph {et~al.}(2015)\citenamefont {Lu},
  \citenamefont {Li}, \citenamefont {Petrov},\ and\ \citenamefont
  {Shlyapnikov}}]{Lu2015a}%
  \BibitemOpen
  \bibfield  {author} {\bibinfo {author} {\bibfnamefont {Z.-K.}\ \bibnamefont
  {Lu}}, \bibinfo {author} {\bibfnamefont {Y.}~\bibnamefont {Li}}, \bibinfo
  {author} {\bibfnamefont {D.~S.}\ \bibnamefont {Petrov}}, \ and\ \bibinfo
  {author} {\bibfnamefont {G.~V.}\ \bibnamefont {Shlyapnikov}},\ }\href
  {\doibase 10.1103/PhysRevLett.115.075303} {\bibfield  {journal} {\bibinfo
  {journal} {Phys. Rev. Lett.}\ }\textbf {\bibinfo {volume} {115}},\ \bibinfo
  {pages} {075303} (\bibinfo {year} {2015})}\BibitemShut {NoStop}%
\bibitem [{\citenamefont {Ferrier-Barbut}\ \emph {et~al.}(2016)\citenamefont
  {Ferrier-Barbut}, \citenamefont {Kadau}, \citenamefont {Schmitt},
  \citenamefont {Wenzel},\ and\ \citenamefont {Pfau}}]{Ferrier-Barbut2016a}%
  \BibitemOpen
  \bibfield  {author} {\bibinfo {author} {\bibfnamefont {I.}~\bibnamefont
  {Ferrier-Barbut}}, \bibinfo {author} {\bibfnamefont {H.}~\bibnamefont
  {Kadau}}, \bibinfo {author} {\bibfnamefont {M.}~\bibnamefont {Schmitt}},
  \bibinfo {author} {\bibfnamefont {M.}~\bibnamefont {Wenzel}}, \ and\ \bibinfo
  {author} {\bibfnamefont {T.}~\bibnamefont {Pfau}},\ }\href {\doibase
  10.1103/PhysRevLett.116.215301} {\bibfield  {journal} {\bibinfo  {journal}
  {Phys. Rev. Lett.}\ }\textbf {\bibinfo {volume} {116}},\ \bibinfo {pages}
  {215301} (\bibinfo {year} {2016})}\BibitemShut {NoStop}%
\bibitem [{\citenamefont {W\"achtler}\ and\ \citenamefont
  {Santos}(2016)}]{Wachtler2016a}%
  \BibitemOpen
  \bibfield  {author} {\bibinfo {author} {\bibfnamefont {F.}~\bibnamefont
  {W\"achtler}}\ and\ \bibinfo {author} {\bibfnamefont {L.}~\bibnamefont
  {Santos}},\ }\href {\doibase 10.1103/PhysRevA.93.061603} {\bibfield
  {journal} {\bibinfo  {journal} {Phys. Rev. A}\ }\textbf {\bibinfo {volume}
  {93}},\ \bibinfo {pages} {061603} (\bibinfo {year} {2016})}\BibitemShut
  {NoStop}%
\bibitem [{\citenamefont {Saito}(2016)}]{Saito2016a}%
  \BibitemOpen
  \bibfield  {author} {\bibinfo {author} {\bibfnamefont {H.}~\bibnamefont
  {Saito}},\ }\href {\doibase 10.7566/JPSJ.85.053001} {\bibfield  {journal}
  {\bibinfo  {journal} {J. Phys. Soc. Jpn}\ }\textbf {\bibinfo {volume} {85}},\
  \bibinfo {pages} {053001} (\bibinfo {year} {2016})}\BibitemShut {NoStop}%
\bibitem [{\citenamefont {Petrov}(2015)}]{Petrov2015a}%
  \BibitemOpen
  \bibfield  {author} {\bibinfo {author} {\bibfnamefont {D.~S.}\ \bibnamefont
  {Petrov}},\ }\href {\doibase 10.1103/PhysRevLett.115.155302} {\bibfield
  {journal} {\bibinfo  {journal} {Phys. Rev. Lett.}\ }\textbf {\bibinfo
  {volume} {115}},\ \bibinfo {pages} {155302} (\bibinfo {year}
  {2015})}\BibitemShut {NoStop}%
\bibitem [{\citenamefont {Lee}\ and\ \citenamefont {Yang}(1957)}]{LY1957}%
  \BibitemOpen
  \bibfield  {author} {\bibinfo {author} {\bibfnamefont {T.~D.}\ \bibnamefont
  {Lee}}\ and\ \bibinfo {author} {\bibfnamefont {C.~N.}\ \bibnamefont {Yang}},\
  }\href {\doibase 10.1103/PhysRev.105.1119} {\bibfield  {journal} {\bibinfo
  {journal} {Phys. Rev.}\ }\textbf {\bibinfo {volume} {105}},\ \bibinfo {pages}
  {1119} (\bibinfo {year} {1957})}\BibitemShut {NoStop}%
\bibitem [{\citenamefont {Lee}\ \emph {et~al.}(1957)\citenamefont {Lee},
  \citenamefont {Huang},\ and\ \citenamefont {Yang}}]{LHY1957}%
  \BibitemOpen
  \bibfield  {author} {\bibinfo {author} {\bibfnamefont {T.~D.}\ \bibnamefont
  {Lee}}, \bibinfo {author} {\bibfnamefont {K.}~\bibnamefont {Huang}}, \ and\
  \bibinfo {author} {\bibfnamefont {C.~N.}\ \bibnamefont {Yang}},\ }\href
  {\doibase 10.1103/PhysRev.106.1135} {\bibfield  {journal} {\bibinfo
  {journal} {Phys. Rev.}\ }\textbf {\bibinfo {volume} {106}},\ \bibinfo {pages}
  {1135} (\bibinfo {year} {1957})}\BibitemShut {NoStop}%
\bibitem [{\citenamefont {Sch\"{u}tzhold}\ \emph {et~al.}(2006)\citenamefont
  {Sch\"{u}tzhold}, \citenamefont {Uhlmann}, \citenamefont {Xu},\ and\
  \citenamefont {Fischer}}]{Schatzhold2006a}%
  \BibitemOpen
  \bibfield  {author} {\bibinfo {author} {\bibfnamefont {R.}~\bibnamefont
  {Sch\"{u}tzhold}}, \bibinfo {author} {\bibfnamefont {M.}~\bibnamefont
  {Uhlmann}}, \bibinfo {author} {\bibfnamefont {Y.}~\bibnamefont {Xu}}, \ and\
  \bibinfo {author} {\bibfnamefont {U.~R.}\ \bibnamefont {Fischer}},\ }\href
  {\doibase 10.1142/S0217979206035631} {\bibfield  {journal} {\bibinfo
  {journal} {Int. J. Mod. Phys. B}\ }\textbf {\bibinfo {volume} {20}},\
  \bibinfo {pages} {3555} (\bibinfo {year} {2006})}\BibitemShut {NoStop}%
\bibitem [{\citenamefont {Lima}\ and\ \citenamefont
  {Pelster}(2011)}]{Lima2011a}%
  \BibitemOpen
  \bibfield  {author} {\bibinfo {author} {\bibfnamefont {A.~R.~P.}\
  \bibnamefont {Lima}}\ and\ \bibinfo {author} {\bibfnamefont {A.}~\bibnamefont
  {Pelster}},\ }\href {\doibase 10.1103/PhysRevA.84.041604} {\bibfield
  {journal} {\bibinfo  {journal} {Phys. Rev. A}\ }\textbf {\bibinfo {volume}
  {84}},\ \bibinfo {pages} {041604} (\bibinfo {year} {2011})}\BibitemShut
  {NoStop}%
\bibitem [{\citenamefont {Lima}\ and\ \citenamefont
  {Pelster}(2012)}]{Lima2012a}%
  \BibitemOpen
  \bibfield  {author} {\bibinfo {author} {\bibfnamefont {A.~R.~P.}\
  \bibnamefont {Lima}}\ and\ \bibinfo {author} {\bibfnamefont {A.}~\bibnamefont
  {Pelster}},\ }\href {\doibase 10.1103/PhysRevA.86.063609} {\bibfield
  {journal} {\bibinfo  {journal} {Phys. Rev. A}\ }\textbf {\bibinfo {volume}
  {86}},\ \bibinfo {pages} {063609} (\bibinfo {year} {2012})}\BibitemShut
  {NoStop}%
\bibitem [{\citenamefont {Altmeyer}\ \emph {et~al.}(2007)\citenamefont
  {Altmeyer}, \citenamefont {Riedl}, \citenamefont {Kohstall}, \citenamefont
  {Wright}, \citenamefont {Geursen}, \citenamefont {Bartenstein}, \citenamefont
  {Chin}, \citenamefont {Denschlag},\ and\ \citenamefont
  {Grimm}}]{Altmeyer2007a}%
  \BibitemOpen
  \bibfield  {author} {\bibinfo {author} {\bibfnamefont {A.}~\bibnamefont
  {Altmeyer}}, \bibinfo {author} {\bibfnamefont {S.}~\bibnamefont {Riedl}},
  \bibinfo {author} {\bibfnamefont {C.}~\bibnamefont {Kohstall}}, \bibinfo
  {author} {\bibfnamefont {M.~J.}\ \bibnamefont {Wright}}, \bibinfo {author}
  {\bibfnamefont {R.}~\bibnamefont {Geursen}}, \bibinfo {author} {\bibfnamefont
  {M.}~\bibnamefont {Bartenstein}}, \bibinfo {author} {\bibfnamefont
  {C.}~\bibnamefont {Chin}}, \bibinfo {author} {\bibfnamefont {J.~H.}\
  \bibnamefont {Denschlag}}, \ and\ \bibinfo {author} {\bibfnamefont
  {R.}~\bibnamefont {Grimm}},\ }\href {\doibase 10.1103/PhysRevLett.98.040401}
  {\bibfield  {journal} {\bibinfo  {journal} {Phys. Rev. Lett.}\ }\textbf
  {\bibinfo {volume} {98}},\ \bibinfo {pages} {040401} (\bibinfo {year}
  {2007})}\BibitemShut {NoStop}%
\bibitem [{\citenamefont {Papp}\ \emph {et~al.}(2008)\citenamefont {Papp},
  \citenamefont {Pino}, \citenamefont {Wild}, \citenamefont {Ronen},
  \citenamefont {Wieman}, \citenamefont {Jin},\ and\ \citenamefont
  {Cornell}}]{Papp2008a}%
  \BibitemOpen
  \bibfield  {author} {\bibinfo {author} {\bibfnamefont {S.~B.}\ \bibnamefont
  {Papp}}, \bibinfo {author} {\bibfnamefont {J.~M.}\ \bibnamefont {Pino}},
  \bibinfo {author} {\bibfnamefont {R.~J.}\ \bibnamefont {Wild}}, \bibinfo
  {author} {\bibfnamefont {S.}~\bibnamefont {Ronen}}, \bibinfo {author}
  {\bibfnamefont {C.~E.}\ \bibnamefont {Wieman}}, \bibinfo {author}
  {\bibfnamefont {D.~S.}\ \bibnamefont {Jin}}, \ and\ \bibinfo {author}
  {\bibfnamefont {E.~A.}\ \bibnamefont {Cornell}},\ }\href {\doibase
  10.1103/PhysRevLett.101.135301} {\bibfield  {journal} {\bibinfo  {journal}
  {Phys. Rev. Lett.}\ }\textbf {\bibinfo {volume} {101}},\ \bibinfo {pages}
  {135301} (\bibinfo {year} {2008})}\BibitemShut {NoStop}%
\bibitem [{\citenamefont {Navon}\ \emph {et~al.}(2011)\citenamefont {Navon},
  \citenamefont {Piatecki}, \citenamefont {G\"unter}, \citenamefont {Rem},
  \citenamefont {Nguyen}, \citenamefont {Chevy}, \citenamefont {Krauth},\ and\
  \citenamefont {Salomon}}]{Navon2011a}%
  \BibitemOpen
  \bibfield  {author} {\bibinfo {author} {\bibfnamefont {N.}~\bibnamefont
  {Navon}}, \bibinfo {author} {\bibfnamefont {S.}~\bibnamefont {Piatecki}},
  \bibinfo {author} {\bibfnamefont {K.}~\bibnamefont {G\"unter}}, \bibinfo
  {author} {\bibfnamefont {B.}~\bibnamefont {Rem}}, \bibinfo {author}
  {\bibfnamefont {T.~C.}\ \bibnamefont {Nguyen}}, \bibinfo {author}
  {\bibfnamefont {F.}~\bibnamefont {Chevy}}, \bibinfo {author} {\bibfnamefont
  {W.}~\bibnamefont {Krauth}}, \ and\ \bibinfo {author} {\bibfnamefont
  {C.}~\bibnamefont {Salomon}},\ }\href {\doibase
  10.1103/PhysRevLett.107.135301} {\bibfield  {journal} {\bibinfo  {journal}
  {Phys. Rev. Lett.}\ }\textbf {\bibinfo {volume} {107}},\ \bibinfo {pages}
  {135301} (\bibinfo {year} {2011})}\BibitemShut {NoStop}%
\bibitem [{\citenamefont {Dalfovo}\ \emph {et~al.}(1999)\citenamefont
  {Dalfovo}, \citenamefont {Giorgini}, \citenamefont {Pitaevskii},\ and\
  \citenamefont {Stringari}}]{Dalfovo1999}%
  \BibitemOpen
  \bibfield  {author} {\bibinfo {author} {\bibfnamefont {F.}~\bibnamefont
  {Dalfovo}}, \bibinfo {author} {\bibfnamefont {S.}~\bibnamefont {Giorgini}},
  \bibinfo {author} {\bibfnamefont {L.}~\bibnamefont {Pitaevskii}}, \ and\
  \bibinfo {author} {\bibfnamefont {S.}~\bibnamefont {Stringari}},\ }\href@noop
  {} {\bibfield  {journal} {\bibinfo  {journal} {Rev.~Mod.~Phys.}\ }\textbf
  {\bibinfo {volume} {71}},\ \bibinfo {pages} {463} (\bibinfo {year}
  {1999})}\BibitemShut {NoStop}%
\bibitem [{Note1()}]{Note1}%
  \BibitemOpen
  \bibinfo {note} {The dipolar LHY local density treatment was formulated in
  \cite {Lima2011a,Lima2012a} and was recently applied as a generalized GPE in
  \cite {Wachtler2016a,Saito2016a}}\BibitemShut {NoStop}%
\bibitem [{\citenamefont {Blakie}\ \emph {et~al.}(2012)\citenamefont {Blakie},
  \citenamefont {Baillie},\ and\ \citenamefont {Bisset}}]{Blakie2012a}%
  \BibitemOpen
  \bibfield  {author} {\bibinfo {author} {\bibfnamefont {P.~B.}\ \bibnamefont
  {Blakie}}, \bibinfo {author} {\bibfnamefont {D.}~\bibnamefont {Baillie}}, \
  and\ \bibinfo {author} {\bibfnamefont {R.~N.}\ \bibnamefont {Bisset}},\
  }\href {\doibase 10.1103/PhysRevA.86.021604} {\bibfield  {journal} {\bibinfo
  {journal} {Phys. Rev. A}\ }\textbf {\bibinfo {volume} {86}},\ \bibinfo
  {pages} {021604} (\bibinfo {year} {2012})}\BibitemShut {NoStop}%
\bibitem [{\citenamefont {Bisset}\ and\ \citenamefont
  {Blakie}(2013)}]{Bisset2013a}%
  \BibitemOpen
  \bibfield  {author} {\bibinfo {author} {\bibfnamefont {R.~N.}\ \bibnamefont
  {Bisset}}\ and\ \bibinfo {author} {\bibfnamefont {P.~B.}\ \bibnamefont
  {Blakie}},\ }\href {\doibase 10.1103/PhysRevLett.110.265302} {\bibfield
  {journal} {\bibinfo  {journal} {Phys. Rev. Lett.}\ }\textbf {\bibinfo
  {volume} {110}},\ \bibinfo {pages} {265302} (\bibinfo {year}
  {2013})}\BibitemShut {NoStop}%
\bibitem [{\citenamefont {Jona-Lasinio}\ \emph {et~al.}(2013)\citenamefont
  {Jona-Lasinio}, \citenamefont {\L{}akomy},\ and\ \citenamefont
  {Santos}}]{JonaLasinio2013}%
  \BibitemOpen
  \bibfield  {author} {\bibinfo {author} {\bibfnamefont {M.}~\bibnamefont
  {Jona-Lasinio}}, \bibinfo {author} {\bibfnamefont {K.}~\bibnamefont
  {\L{}akomy}}, \ and\ \bibinfo {author} {\bibfnamefont {L.}~\bibnamefont
  {Santos}},\ }\href {\doibase 10.1103/PhysRevA.88.013619} {\bibfield
  {journal} {\bibinfo  {journal} {Phys. Rev. A}\ }\textbf {\bibinfo {volume}
  {88}},\ \bibinfo {pages} {013619} (\bibinfo {year} {2013})}\BibitemShut
  {NoStop}%
\bibitem [{\citenamefont {Hutchinson}\ \emph {et~al.}(1998)\citenamefont
  {Hutchinson}, \citenamefont {Dodd},\ and\ \citenamefont
  {Burnett}}]{Hutchinson1998a}%
  \BibitemOpen
  \bibfield  {author} {\bibinfo {author} {\bibfnamefont {D.~A.~W.}\
  \bibnamefont {Hutchinson}}, \bibinfo {author} {\bibfnamefont {R.~J.}\
  \bibnamefont {Dodd}}, \ and\ \bibinfo {author} {\bibfnamefont
  {K.}~\bibnamefont {Burnett}},\ }\href@noop {} {\bibfield  {journal} {\bibinfo
   {journal} {Phys. Rev. Lett.}\ }\textbf {\bibinfo {volume} {81}},\ \bibinfo
  {pages} {2198} (\bibinfo {year} {1998})}\BibitemShut {NoStop}%
\bibitem [{\citenamefont {Castin}\ and\ \citenamefont
  {Dum}(1998)}]{Castin1998a}%
  \BibitemOpen
  \bibfield  {author} {\bibinfo {author} {\bibfnamefont {Y.}~\bibnamefont
  {Castin}}\ and\ \bibinfo {author} {\bibfnamefont {R.}~\bibnamefont {Dum}},\
  }\href {\doibase 10.1103/PhysRevA.57.3008} {\bibfield  {journal} {\bibinfo
  {journal} {Phys. Rev. A}\ }\textbf {\bibinfo {volume} {57}},\ \bibinfo
  {pages} {3008} (\bibinfo {year} {1998})}\BibitemShut {NoStop}%
\bibitem [{\citenamefont {Morgan}(2000)}]{Morgan2000a}%
  \BibitemOpen
  \bibfield  {author} {\bibinfo {author} {\bibfnamefont {S.~A.}\ \bibnamefont
  {Morgan}},\ }\href@noop {} {\bibfield  {journal} {\bibinfo  {journal} {J.
  Phys. B}\ }\textbf {\bibinfo {volume} {33}},\ \bibinfo {pages} {3847}
  (\bibinfo {year} {2000})}\BibitemShut {NoStop}%
\bibitem [{\citenamefont {Morgan}(2004)}]{Morgan2004a}%
  \BibitemOpen
  \bibfield  {author} {\bibinfo {author} {\bibfnamefont {S.~A.}\ \bibnamefont
  {Morgan}},\ }\href {\doibase 10.1103/PhysRevA.69.023609} {\bibfield
  {journal} {\bibinfo  {journal} {Phys. Rev. A}\ }\textbf {\bibinfo {volume}
  {69}},\ \bibinfo {pages} {023609} (\bibinfo {year} {2004})}\BibitemShut
  {NoStop}%
\bibitem [{\citenamefont {Cormack}\ and\ \citenamefont
  {Hutchinson}(2012)}]{Cormack2012a}%
  \BibitemOpen
  \bibfield  {author} {\bibinfo {author} {\bibfnamefont {S.~C.}\ \bibnamefont
  {Cormack}}\ and\ \bibinfo {author} {\bibfnamefont {D.~A.~W.}\ \bibnamefont
  {Hutchinson}},\ }\href {\doibase 10.1103/PhysRevA.86.053619} {\bibfield
  {journal} {\bibinfo  {journal} {Phys. Rev. A}\ }\textbf {\bibinfo {volume}
  {86}},\ \bibinfo {pages} {053619} (\bibinfo {year} {2012})}\BibitemShut
  {NoStop}%
\bibitem [{\citenamefont {Ticknor}(2012)}]{Ticknor2012a}%
  \BibitemOpen
  \bibfield  {author} {\bibinfo {author} {\bibfnamefont {C.}~\bibnamefont
  {Ticknor}},\ }\href {\doibase 10.1103/PhysRevA.85.033629} {\bibfield
  {journal} {\bibinfo  {journal} {Phys. Rev. A}\ }\textbf {\bibinfo {volume}
  {85}},\ \bibinfo {pages} {033629} (\bibinfo {year} {2012})}\BibitemShut
  {NoStop}%
\bibitem [{\citenamefont {Eberlein}\ \emph {et~al.}(2005)\citenamefont
  {Eberlein}, \citenamefont {Giovanazzi},\ and\ \citenamefont
  {O'Dell}}]{Eberlein2005a}%
  \BibitemOpen
  \bibfield  {author} {\bibinfo {author} {\bibfnamefont {C.}~\bibnamefont
  {Eberlein}}, \bibinfo {author} {\bibfnamefont {S.}~\bibnamefont
  {Giovanazzi}}, \ and\ \bibinfo {author} {\bibfnamefont {D.~H.~J.}\
  \bibnamefont {O'Dell}},\ }\href {\doibase 10.1103/PhysRevA.71.033618}
  {\bibfield  {journal} {\bibinfo  {journal} {Phys. Rev. A}\ }\textbf {\bibinfo
  {volume} {71}},\ \bibinfo {pages} {033618} (\bibinfo {year}
  {2005})}\BibitemShut {NoStop}%
\bibitem [{Note2()}]{Note2}%
  \BibitemOpen
  \bibinfo {note} {In some regimes the full GPE solutions can have peak density
  occurring away from trap center (e.g.~see \cite
  {Ronen2006a,Lu2010a,Martin2012a}).}\BibitemShut {Stop}%
\bibitem [{Note3()}]{Note3}%
  \BibitemOpen
  \bibinfo {note} {For the case of $a_s/a_0=70$ we use $\kappa =1/20$, and for
  $a_s/a_0=100$ we use $\kappa =1/5$ as estimates of the droplet aspect ratios
  based on our results in Fig.~\ref {Fig:Dy_N} (a) and (b)}\BibitemShut
  {NoStop}%
\bibitem [{Note4()}]{Note4}%
  \BibitemOpen
  \bibinfo {note} {The last term of Eq.~(29) of \cite {Lima2012a} is too high
  by a factor of $2$}\BibitemShut {NoStop}%
\bibitem [{Note5()}]{Note5}%
  \BibitemOpen
  \bibinfo {note} {This is consistent with Eq.~5 of \cite
  {Wachtler2016a}, except that their $-q_c(\theta )^2/2$ should be $-q_c(\theta
  )^2/10$.}\BibitemShut {Stop}%
\bibitem [{\citenamefont {Martin}\ and\ \citenamefont
  {Blakie}(2012)}]{Martin2012a}%
  \BibitemOpen
  \bibfield  {author} {\bibinfo {author} {\bibfnamefont {A.~D.}\ \bibnamefont
  {Martin}}\ and\ \bibinfo {author} {\bibfnamefont {P.~B.}\ \bibnamefont
  {Blakie}},\ }\href {\doibase 10.1103/PhysRevA.86.053623} {\bibfield
  {journal} {\bibinfo  {journal} {Phys. Rev. A}\ }\textbf {\bibinfo {volume}
  {86}},\ \bibinfo {pages} {053623} (\bibinfo {year} {2012})}\BibitemShut
  {NoStop}%
\end{thebibliography}

%

\end{document}